\documentclass[amsmath,amssymb,aps,pre,groupedaddress,superscriptaddress,twocolumn]{revtex4-2}
\usepackage[utf8]{inputenc}
\usepackage[english]{babel}
\usepackage[english]{varioref}
\usepackage{graphicx}
\usepackage{dcolumn}
\usepackage{bm}
\usepackage{amsmath}
\usepackage{amssymb}
\usepackage{xcolor}
\usepackage[font=small]{caption}
\usepackage{mathrsfs}
\usepackage{tabularx}
\usepackage[colorlinks=true,linkcolor=blue,urlcolor=blue,citecolor=blue,anchorcolor=blue]{hyperref}

\begin{document}

\title{Adaptive behaviors neutralize bistable explosive transitions in higher-order contagion}

\author{Marco Mancastroppa}
%\email{marco.mancastroppa@cpt.univ-mrs.fr}
\affiliation{Aix Marseille Univ, Université de Toulon, CNRS, CPT, Turing Center for Living Systems, 13009 Marseille, France}
\author{Márton Karsai}
\affiliation{Department of Network and Data Science, Central European University, 1100 Vienna, Austria}
\affiliation{National Laboratory for Health Security, HUN-REN Rényi Institute of Mathematics, 1053 Budapest, Hungary}
\author{Alain Barrat}
\affiliation{Aix Marseille Univ, Université de Toulon, CNRS, CPT, Turing Center for Living Systems, 13009 Marseille, France}

\begin{abstract}
During contagion phenomena, individuals perceiving a risk of infection commonly adapt their behavior and reduce their exposure. The effects of such adaptive mechanisms have been studied for processes in which pairwise interactions drive contagion. However, contagion  
and the perception of infection risk  can also involve ("higher-order") group interactions, leading potentially to new phenomenology.
%can also involve ("higher-order") group interactions, leading e.g. to the emergence of discontinuous phase transitions. 
How adaptive behavior resulting from risk perception 
affects higher-order processes remains an open question. Here, we consider the impact of several risk-based adaptive behaviors 
on pairwise and higher-order contagion processes, using numerical simulations and an analytical mean-field approach. For pairwise contagion, adaptive mechanisms based on local (pairwise or group-based) risk perception impact only the endemic state, without affecting the epidemic phase transition. For higher-order contagion processes, instead, the adaptivity defuses the impact of non-linear group interactions: this reduces or even completely suppresses the parameter range in which bistability is possible, effectively transforming 
a higher-order contagion process into a pairwise one.
\end{abstract}

\maketitle

Many complex phenomena, such as the spread of epidemics, 
information and opinions, can be represented by contagion processes on networks \cite{barrat2008dynamical,satorras2015epidemic,majhi2022,WANG20241}. 
The description of the underlying infection mechanisms can involve either
pairwise interactions (as in the modelling of epidemic spread, in which  each
transmission event corresponds to a single exposure to an infectious individual \cite{barrat2008dynamical,satorras2015epidemic}) or 
(higher-order) group interactions. The latter description is often used for 
social contagion processes, where non-linear reinforcement effects can occur in groups
\cite{Battiston2021,BATTISTON20201,majhi2022,WANG20241}. Recent research
has shown how these higher-order interactions give rise to a rich phenomenology, including discontinuous transitions, multi-stability regimes, critical mass effects
\cite{Iacopini2019,bick2021,Battiston2021,FerrazdeArruda2023,FerrazdeArruda2021,WANG20241,Kiss2025,Malizia2025,malizia2025disentangling,St-Onge2022,iacopini2022group,Millan2025},
and contagion paths differing from the ones due to pairwise interactions \cite{Mancastroppa2023,cencetti2023,andres2025}.

While the structure of interactions strongly impacts the unfolding of a contagion dynamics at both microscopic and macroscopic levels, the contagion process can also itself reshape the interactions, if individuals are aware of the contagion risk and adapt their behavior accordingly
\cite{gross2009adaptive,berner2023adaptive,gross2008,SAYAMA20131645,funk2010review,perra2011adaptive,verelst2016behavioural,de2020relationships,Degaetano2024modeling}. 
For instance, if a disease is spreading in a population, individuals can become
aware of the spread through information available either globally at the 
population level \cite{berner2023adaptive,gross2008,perra2011adaptive}, or
locally as other individuals become infectious in their neighborhood
\cite{gross2006adaptive,bagnoli2007risk,marceau2010,moinet2018effect}.
Both healthy and infectious individuals may then modify their behavior to avoid being contaminated or infecting others 
\cite{feniche2011adaptive,funk2009awearness,Ferguson2007,odor2025}. These 
adaptive behaviors can consist either in the removal or rewiring of interactions, or in the implementation of protective measures (masks, handwashing, etc) that decrease the probability of infection \cite{berner2023adaptive,gross2008,perra2011adaptive}. 

In the context of pairwise contagion processes, many models have been proposed
and studied to explore the effect of such adaptative mechanisms based
on awareness, obtained also through pairwise interactions 
\cite{gross2006adaptive,bagnoli2007risk,marceau2010,guo2013,karsai2025,odor2025,moinet2018effect,mancastroppa2024,scarpino2016effect,mancastroppa2020active,Mancastroppa2021CT}.
However, the impact of behavior change on higher-order contagion processes,
including adaptive mechanisms triggered by group-driven information,
remains largely unexplored \cite{Burgio2025,liu2025higher,mancastroppa2025,lucas2024simplicially}.

Here and in a companion paper \cite{mancastroppa2025}, we contribute to filling this gap by considering both pairwise and non-linear higher-order contagion processes in the presence of adaptive strategies based on local information, driven either by pairwise or by group information. 
We focus in this paper on the impact of these strategies on the epidemic phase 
transition. To this aim, we consider two specific mechanisms, where the adaptivity is driven by either the number of infectious neighbors of an individual (pairwise-based 
mechanism) or by the number of infectious groups to which an individual
belongs (group-based mechanism).
%In \cite{mancastroppa2025}, we explore a range of pairwise and group-based adaptive mechanisms and provide a systematic investigation of their relative effectivity and cost. Here, we focus on their impact on the epidemic phase transition. We consider two specific mechanisms (results are extended to other mechanisms in \cite{mancastroppa2025}), where the adaptivity is driven by either the number of infectious neighbors of an individual (pairwise-based  mechanism) or by the number of infectious groups to which an individual belongs (group-based mechanism).
Through numerical simulations and a mean-field analytical approach, we show that in pairwise contagion processes, these adaptive mechanisms impact the endemic state but do not influence the nature nor the location of the epidemic phase transition, which remains continuous. 
For higher-order contagion processes however, the adaptive interaction change has a strong impact on the epidemic transition itself: we show that it defuses the non-linearity and explosiveness of group contagion events, which are responsible for the emergence of a discontinuous transition and of a bistable regime in the non-adaptive case \cite{St-Onge2022,FerrazdeArruda2023}.
As a result, adaptive behaviors shrink the bistability region and decrease the discontinuity of the phase transition, possibly even completely suppressing it, reverting the transition to a continuous one as in pairwise contagion.
In \cite{mancastroppa2025}, we extend the results to a wide 
range of pairwise and group-based adaptive mechanisms, providing a systematic investigation of their relative effectivity and cost.

\begin{figure*}[ht!]
\includegraphics[width=\textwidth]{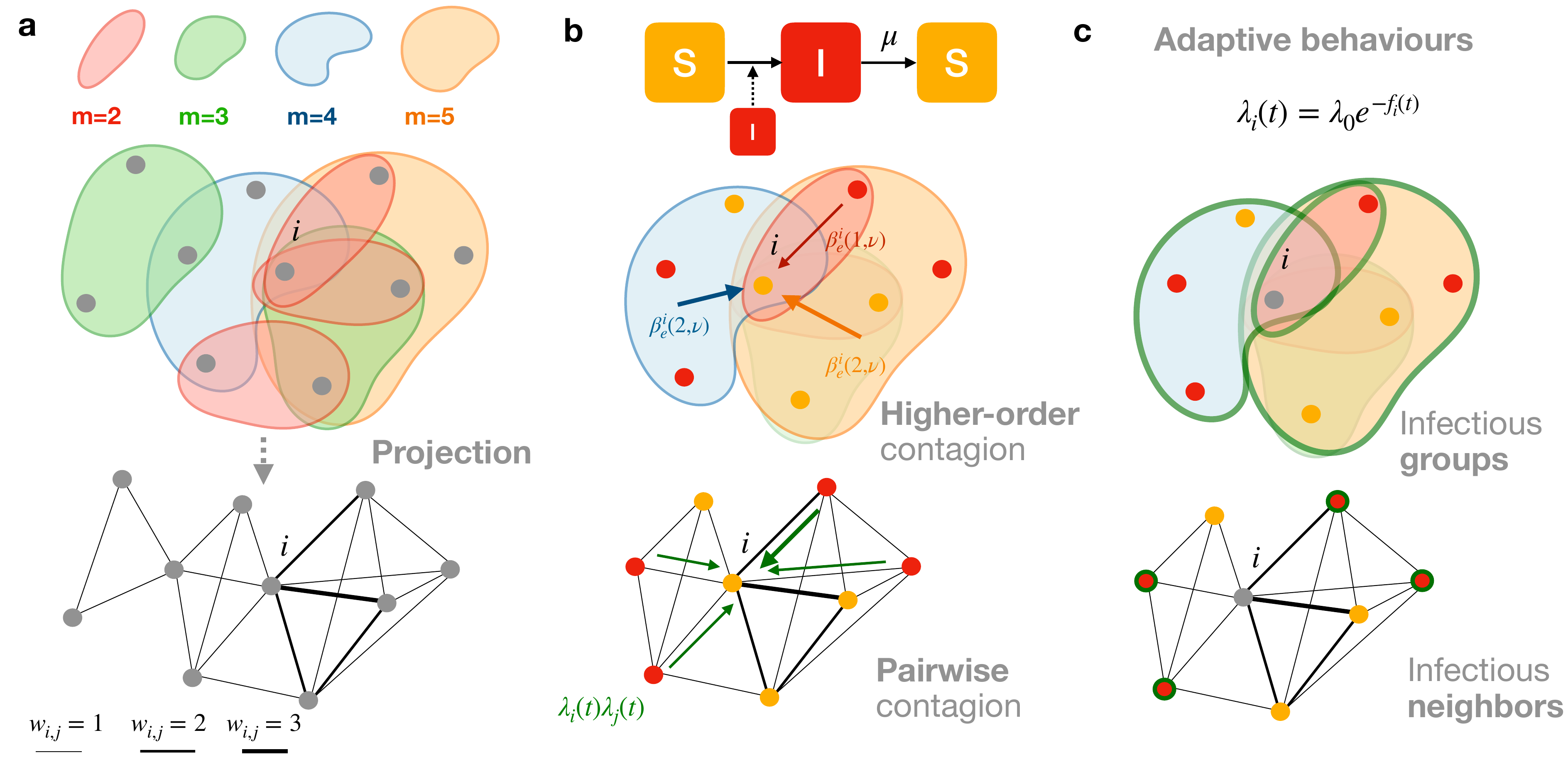}
\caption{\textbf{Schematic representation of contagion and adaptation mechanisms.} \textbf{a}: example of a hypergraph $\mathcal{H}$ and the corresponding weighted projected graph $\mathcal{G}$. \textbf{b}: the SIS contagion mechanisms for the node $i$, with pairwise (bottom) and higher-order (top) processes. 
\textbf{c}: adaptive strategies for node $i$, with group-based awareness (top, $ng$ with $\theta=0.3$: 
$f_i^{ng}(t)=3/\langle D \rangle$) or pairwise-based awareness (bottom, $nn$: $f_i^{nn}(t)=4/\langle k \rangle  $).}
\label{fig:figure1}
\end{figure*}

\textit{Contagion models.} We consider a hypergraph $\mathcal{H}=(\mathcal{V},\mathcal{E})$, where $\mathcal{V}$ denotes the set of its nodes, $\mathcal{E}$ is the set of its hyperedges, and its weighted projected graph is $\mathcal{G}=(\mathcal{V},\mathcal{L})$ (see Fig.~\ref{fig:figure1}\textbf{a} and Appendix). 
These structures constitute the substrate for the spread of a 
Susceptible-Infectious-Susceptible (SIS) model for epidemic spread in discrete time.
Awareness of the spread (which drives adaptive mechanisms) is also sustained by $\mathcal{H}$ and $\mathcal{G}$.
In the SIS model, each individual is at each time step in one of two states: susceptible
(S) individuals can become infectious (I) through contagion events, i.e., interactions
with infectious individual(s), and infectious individuals recover spontaneously, with probability $\mu$ per time step. 
To describe in a phenomenological way both the risk of infection and of transmission of the nodes, each node $i \in \mathcal{V}$ is assigned a parameter $\lambda_i(t)$. The probability for a susceptible node $i$ to become infectious through interactions with a set of infectious nodes $j$ in its neighborhood $\mathcal{N}(i)$ depends on $\lambda_i(t)$ and on $\{\lambda_j(t)\}_{j \in  \mathcal{N}(i) \cap  \mathcal{I}(t)}$ (where  $\mathcal{I}(t)$ is the set of infectious nodes at time $t$). The parameters $\lambda_i(t)$ can moreover be modulated in time to encode the effect of the nodes' awareness of the epidemic.

We consider two different contagion mechanisms (Fig.~\ref{fig:figure1}\textbf{b}), based either on pairwise transmission or on higher-order non-linear effects. In the \textit{pairwise contagion process}, a susceptible node $i$ at time $t$ has probability $\lambda_i(t) \lambda_j(t)$ to get infected by each infectious node $j$ in $\mathcal{N}(i)$ (each link $l \in \mathcal{G}$ can thus be source of infection with multiplicity equal to its weight $w_l$).
In the \textit{higher-order contagion process}, a susceptible node $i$ belonging to a hyperedge $e \in \mathcal{E}$, with $i_e$ infected individuals at time $t$, is infected with probability 
$\beta_e^i(i_e,\nu)=\lambda_0 i_e^{\nu} \lambda_i(t) \langle \lambda_j(t)/\lambda_0 \rangle_{i_e}^{\nu}$. Here $\langle \cdot \rangle_{i_e} $ indicates the average over the $i_e$ infected nodes in $e$ at time $t$ and $\nu$ sets the non-linearity of the process, modelling reinforcement group effects \cite{St-Onge2022}. 
%Note that the non-adaptive case with $\lambda_j(t) = \lambda_0 \forall t, \forall j \in \mathcal{V}$ corresponds to the non-linear higher-order contagion model studied in \cite{St-Onge2022}.

\textit{Adaptive mechanisms.} We consider adaptive behaviors driven by risk perception due to local information (epidemic awareness) conveyed through the same interactions along which contagion can occur. We assume that these behaviors do not change the hypergraph topology, but act on the parameters $\lambda_i(t)$ of both susceptible and infectious nodes.
This mechanism models a mild adaptation, where agents increase their attention during social interactions (e.g., through physical distance, use of masks, and increased hygiene), without changing the social structure but resulting in both a self-protective effect (for susceptible nodes) and an altruistic effect (for infectious ones) \cite{karsai2025}.

The baseline processes are represented by the \textit{non-adaptive (NAD)} case with $\lambda_i(t)=\lambda_0 \forall i \in \mathcal{V}, \forall t$, corresponding to the standard pairwise or higher-order SIS processes as in \cite{St-Onge2022}.
For the adaptive behaviors, we assume an exponential reduction $\lambda_i(t)= \lambda_0 e^{-f_i(t)}$~\cite{bagnoli2007risk,karsai2025,moinet2018effect}, where $f_i(t)$ encodes the awareness of node $i$. Its specific form 
%is independent of the type of contagion process considered and 
defines the adaptation strategy considered, i.e., how the node's risk 
perception depends on the health status of its neighborhood at time $t$.
We consider here two such strategies (see Fig. \ref{fig:figure1}\textbf{c}). 
In the \textit{``infectious neighbors''} ($nn$) strategy, 
the awareness of a node is based on pairwise information, coming from its neighbors on the projected network $\mathcal{G}$:
\begin{equation}
    f_i^{nn}(t) = |\mathcal{N}(i) \cap  \mathcal{I}(t)| /\langle k \rangle,
\end{equation}
where $\langle k \rangle$ is the average degree in $\mathcal{G}$
\footnote{In \cite{mancastroppa2025} we also consider a version in which each infectious neighbor $j$ of $i$ contributes to the awareness proportionally to the weight of the link $ij$.}.
In the \textit{``infectious groups''} ($ng$) strategy instead, 
the awareness of a node depends on the number of ``infectious'' hyperedges (groups) in which it participates: a group is defined as infectious for $i$ if the number $i_e$ of infected individuals within it {\it perceived by $i$} (i.e., distinct from $i$) is $i_e> \theta (|e|-1)$, where $\theta$ is a fixed threshold (parameter of the strategy). 
In this case:
\begin{equation}
    f_i^{ng}(t) = \sum\limits_{e \in \mathcal{E}(i)} \bm{1}_{\theta}(e,i)/\langle D \rangle ,
\end{equation}
where $\langle D \rangle$ is the average hyperdegree in $\mathcal{H}$
(i.e., the average number of hyperedges of a node), $\mathcal{E}(i)$ is the set of hyperedges involving $i$ and $\bm{1}_{\theta}(e,i)$ is the group-indicator function, equal to $1$ if $i_e> \theta (|e|-1)$ (without counting $i$) and $0$ otherwise. The parameter $\theta \in [0,1]$ sets the level of alert: hereafter we will consider $\theta=0.3$, while in \cite{mancastroppa2025} we investigate other values of $\theta$.
In \cite{mancastroppa2025}, we also consider strategies involving mixed pairwise and higher-order information, or based on the fraction of infectious neighbors or groups rather than on their numbers.

\begin{figure*}[ht!]
\includegraphics[width=\textwidth]{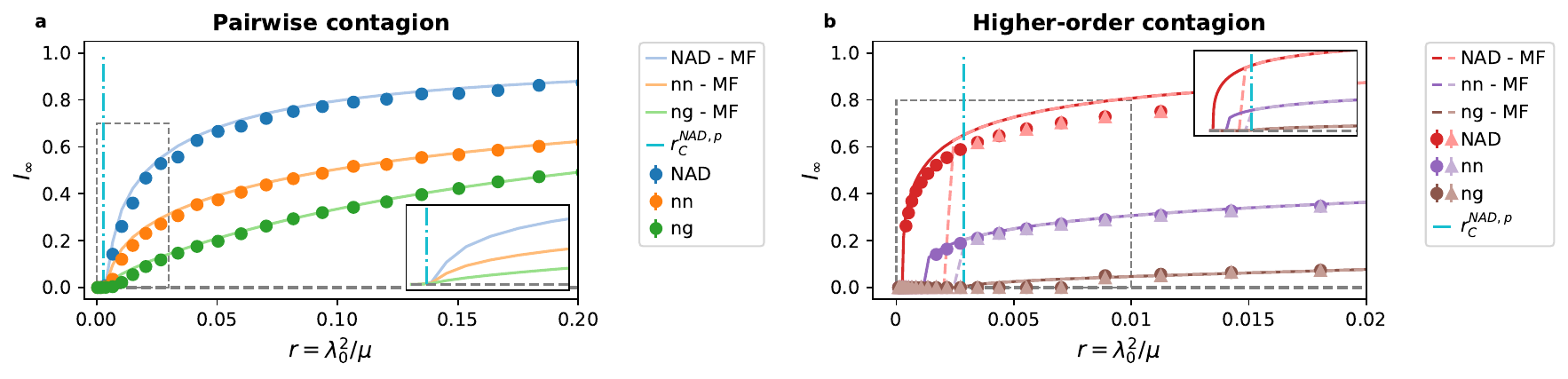}
\caption{\textbf{Phase diagram for pairwise and higher-order contagion processes.} \textbf{a}, \textbf{b}: the epidemic prevalence in the asymptotic steady state, $I_{\infty}$, as a function of the effective infection rate $r$, for the non-adaptive case (NAD) and for the two adaptation strategies ($nn$, $ng$), respectively for the pairwise and higher-order contagion process. 
Lines give the results of the numerical integration of the IBMF equations, and markers show the results of numerical simulations. 
For the higher-order processes, we consider simulations and numerical integration starting either with a low (triangles and dashed lines) or with a high (dots and solid lines) fraction of infected nodes. 
We consider as substrate a dataset describing interactions in a hospital and the numerical results are averaged over $300$ simulations. Here $\nu=4$, $\theta=0.3$. The insets represent a zoom on the mean-field results in the area marked by a dashed rectangle, and the vertical line indicates $r_C^{NAD,p}$.
Note that the error-bars on numerical simulations are smaller than the corresponding markers (considering one standard deviation).
}
\label{fig:figure2}
\end{figure*}

\textit{Mean-field approach.} To analyze the coupling between contagion dynamics and adaptive behaviors, we rely on an individual-based mean-field approach (IBMF), which describes the temporal evolution of the  probability $P_i(t)$ that node $i$ is infectious at time $t$.
IBMF neglects dynamical correlations between the states of neighboring nodes but takes into account the actual structure of interactions (i.e., the graph or hypergraph on which contagion processes occur) \cite{satorras2015epidemic}. 
For the pairwise contagion, $P_i(t)$ evolves according to:
\begin{equation}
    \partial_t P_i(t)= -\mu P_i(t) + (1-P_i(t)) \sum\limits_{j \in \mathcal{V}}  w_{i,j} \lambda_i(t) \lambda_j(t) P_j(t),
    \label{eq:MF_p}
\end{equation}
where the first term accounts for the spontaneous recovery process and the second term describes the contagion of $i$, when it is susceptible, by one of its infectious neighbors.

For higher-order contagion, $P_i(t)$ evolves according to:
\begin{widetext}
\begin{equation}
    \partial_t P_i(t)= -\mu P_i(t) + (1-P_i(t)) \sum\limits_{e \in \mathcal{E}(i)} 
    \left[ 
    \sum\limits_{i_e=1}^{|e|-1} Q_{e \setminus i}(i_e,t) 
    \lambda_0 i_e^\nu \lambda_i(t) 
    \overline{\langle \lambda_j(t)/\lambda_0 \rangle_{i_e}^\nu}
    \right],
    \label{eq:MF_HO}
\end{equation}
\end{widetext}
where the second term on the rhs. takes into account the probability that $i$ is infected through a hyperedge in which it participates. In particular, $Q_{e \setminus i}(i_e,t) $ is the probability that in the hyperedge $e$ at time $t$ there are exactly $i_e$ infectious individuals (distinct from $i$); $\overline{\langle \lambda_j(t)/\lambda_0 \rangle_{i_e}^\nu}$ denotes the average over these $i_e$ infectious individuals 
and over all possible configurations of $i_e$ infectious individuals in $e \backslash i$. Equations \eqref{eq:MF_p} and Eq.s \eqref{eq:MF_HO} constitute two closed sets of $N=|\mathcal{V}|$ coupled non-linear differential equations. Indeed, we show in
the Appendix how $\lambda_i(t)$, $Q_{e \setminus i}(i_e,t) $ and $\overline{\langle \lambda_i(t)/\lambda_0 \rangle_{i_e}^\nu}$ can be analytically obtained 
from the probabilities $P_j(t)$ of all nodes $j$.

\textit{Results.} 
We perform direct Monte-Carlo numerical simulations of the pairwise and higher-order contagion
processes in discrete time, as well as a numerical integration of the mean-field equations, obtaining in each case the average epidemic prevalence in the asymptotic stationary state $I_\infty \equiv |\mathcal{I}|(t \to \infty)$ (see Appendix). 
As substrate for the spread, we consider several empirical hypergraphs 
describing human face-to-face interactions in different settings \cite{sociopatterns,Vanhems2013,genois2018,Genois2023,Stehle2011,Mastrandrea2015},
and also synthetic hypergraphs (see Appendix and Supplementary Material, SM). 
Here we present results for a dataset describing interactions in a hospital \cite{sociopatterns,Vanhems2013}, and we refer to the SM for similar results on other hypergraphs. Figure \ref{fig:figure2} displays $I_\infty$ as a function of the reduced infection rate $r=\lambda_0^2/\mu$ for pairwise and higher-order processes in the non-adaptive case as well as for the two adaptive strategies considered: the IBMF framework and the numerical simulations show a very good agreement in all cases.

We refer to \cite{mancastroppa2025} for a detailed investigation of the ability of a wide set of adaptive strategies (including $ng$ and $nn$) to reduce the asymptotic epidemic prevalence of pairwise and higher-order processes in the active phase 
of the epidemics, as well as a study of the associated societal costs (in terms
of reduction of interactions). In Fig.~\ref{fig:figure2} we simply highlight how both $nn$ and $ng$ strategies drastically reduce $I_\infty$ for pairwise and higher-order processes, with a stronger effect for the group-based awareness ($ng$).
Here, we focus on a more fundamental aspect, namely, how adaptive strategies impact the epidemic phase transition separating the disease-free absorbing phase 
($I_\infty =0$) from the active phase with $I_\infty > 0$ \cite{satorras2015epidemic}.

In the pairwise contagion process, the non-adaptive case presents the well-known continuous phase transition \cite{satorras2015epidemic}. 
Interestingly, the adaptive mechanisms do not alter the phase transition, which remains continuous and presents the same epidemic threshold (see Fig. \ref{fig:figure2}\textbf{a}). We show in the Appendix that it is actually possible to obtain analytically the equality between the epidemic threshold of adaptive and non-adaptive case:
\begin{equation}
    r_C^{NAD,p}=r_C^{nn,p}=r_C^{ng,p}=1/\Lambda_w,
    \label{eq:rcp}
\end{equation}
where $\Lambda_w$ is the maximum eigenvalue of the weights matrix $\{w_{i,j}\}$. 
This can be understood intuitively quite simply. Indeed, as the epidemic transition
is continuous, the fraction of infected nodes tends to $0$ in the active phase close
to the critical point. Thus, $f_i^{nn}, f_i^{ng} \ll 1$ and $\lambda_i \sim \lambda_0 \forall i$ for $r$ close to $r_C^{NAD,p}$: the effect of the adaptation strategies 
becomes of second order and cannot shift the transition point.

The higher-order contagion process presents a different and more interesting
phenomenology. Indeed, the non-adaptive case presents a discontinuous phase transition, with a bistability region where both the endemic and absorbing states are stable \cite{St-Onge2022}. 
However, as shown in Fig. \ref{fig:figure2}\textbf{b}, both adaptive strategies shrink the bistability region. Specifically, the $nn$ strategy increases the value of the lower bound of this region. Strikingly, the $ng$ strategy completely neutralizes the bistability regime and the explosiveness of the transition, making it continuous with the same epidemic threshold as the pairwise contagion, $r_C^{NAD,p}$.

Although the adaptive strategies are based on the same mechanisms in pairwise and
higher-order contagion, they have drastically different effects.
This can be understood both intuitively and at a more fundamental level as follows.
As the transition in the higher-order case is discontinuous, the prevalence in the NAD case takes a large value right above the transition point. Therefore, $f_i^{nn}$ and $f_i^{ng}$ are not necessarily small as in the pairwise case. Thus $\lambda_i$ can become locally much smaller than $\lambda_0$, bringing the prevalence down and making the
high-prevalence state unstable for $r$ not large enough.

\begin{figure*}[ht!]
\includegraphics[width=\textwidth]{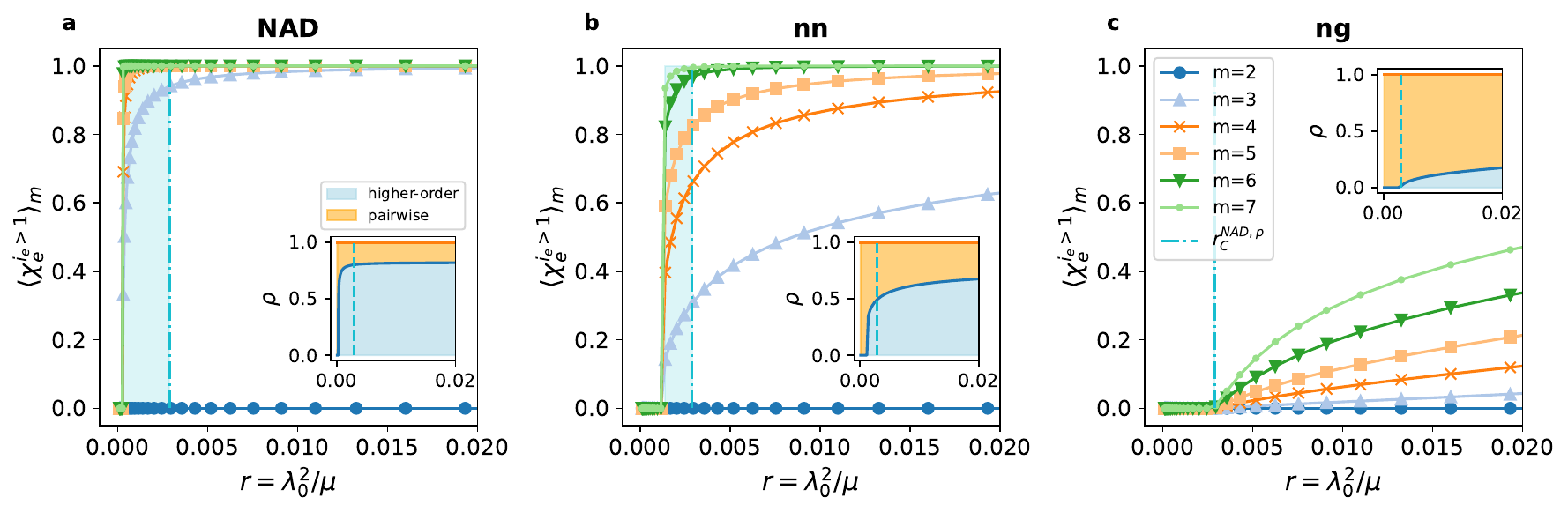}
\caption{\textbf{Neutralization of higher-order contagion.} 
Probability $\langle \chi_e^{i_e>1} \rangle_m$ that a group of size $m$ 
with at least one susceptible node has more than $1$ infected individuals in the asymptotic steady state, as a function of the effective infection rate $r$, 
for different values of $m$. 
$\langle \chi_e^{i_e>1} \rangle_m$ is estimated through numerical integration of the mean-field equations, with initial conditions having high fraction of infected nodes. 
Panel \textbf{a} corresponds to the non-adaptive case, and panels \textbf{b,c} each to an adaptive strategy. We use $\nu=4$ and $\theta=0.3$, the vertical line indicates $r_C^{NAD,p}$ and the light-blue colored area indicates the bistability region. 
The inset shows the probability $\rho$ that a group of arbitrary size supports higher-order contagion (light-blue area), i.e., a contagion event triggered by more than one infectious, as a function of $r$, and the complementary probability that it supports pairwise contagion (orange area), see Appendix for how to compute $\rho$.
}
\label{fig:figure3}
\end{figure*}

We can investigate this point in a more quantitative manner by recalling that the discontinuity of the transition and the bistability regime are due to the non-linearity in the contagion probability \cite{St-Onge2022}, which are relevant only in 
hyperedges where more than one infectious node is present.
We thus consider the probability (in the stationary state) $\chi_e^{i_e>1}$ that a hyperedge $e$ with at least one susceptible individual (i.e. where a contagion can occur) has more than two infectious individuals. We show in the Appendix how this probability can be computed from the $P_i$ in the IBMF framework and we present the results in Fig. \ref{fig:figure3}. 
In the NAD case, the average of this quantity over hyperedges of size $m$ obeys
$\langle \chi_e^{i_e>1} \rangle_m \sim 1$ for all $m>2$ in the bistability region. The contagion events involving the non-linear higher-order mechanism are thus sustained in this region. In fact, the propagation is predominantly driven by group contagions, %while pairwise contagion is limited, 
as measured by the fraction $\rho$ of potential infections driven by non-linear group processes (see Appendix and Fig. \ref{fig:figure3}\textbf{a}).

Adaptive behaviors reduce the probability of contagion within groups in which the infectious nodes are located, by locally increasing awareness (see SM). Figure \ref{fig:figure3}\textbf{b} and \textbf{c} show how both strategies lead to a strong decrease of $\chi_e^{i_e>1}$ for all hyperedge sizes, leading to a reduction of the fraction $\rho$ of contagions driven by non-linear group mechanisms. Higher-order effects are thus suppressed by the strategies, which make it more difficult to have several infectious nodes at the same time in the same hyperedge, replacing group contagion events by pairwise ones. As a result, the discontinuous transition at the lower bound of the bistability region is pushed towards higher values of $r$, and the amplitude of the discontinuity is reduced. Instead, the upper bound of the bistability region, which corresponds to the pairwise epidemic threshold, is not modified, as the stability of the absorbing state with $I_\infty = 0$ is determined solely by pairwise interactions (see Appendix) \cite{FerrazdeArruda2021}. 
As a result, the width of the bistability region decreases. For adaptive strategies having a mild effect, such as the $nn$ one, $\chi_e^{i_e>1}$ and $\rho$ are decreased for all $r$ but go to zero only for $r$ small enough: the bistability region shrinks but the transition remains discontinuous (Fig. \ref{fig:figure3}\textbf{b}).
On the other hand, for adaptive strategies efficient enough to make $\chi_e^{i_e>1} = 0 \, \forall e$ for all $r < r_C^{NAD,p}$, such as the $ng$ strategy, only pairwise contagions can be sustained in this regime (Fig. \ref{fig:figure3} \textbf{c}). As a result, since only the absorbing state is stable for pairwise contagion when $r < r_C^{NAD,p}$, the bistability region completely disappears and the phase transition changes from discontinuous to continuous.
This can also be shown analytically at the mean-field level: if we assume that $\chi_e^{i_e>1} \sim 0$, the mean-field equations for higher-order contagion reduce to those of pairwise contagion around the critical point (see Appendix), thus leading to a continuous transition at the same epidemic threshold. 
Note that these considerations apply only for $r \leq r_C^{NAD,p}$: for $r > r_C^{NAD,p}$ the pairwise contagion by itself leads to a non-zero $I_\infty$, and the prevalence increases with $r$, so that $\chi_e^{i_e>1}$ also increases and non-linear group contagion events become sustainable again, hence altering the dynamics compared to the pairwise case.

We show in the SM that these findings are robust for processes unfolding over other empirical and synthetic hypergraphs, and we highlight the role of the underlying structure \cite{mancastroppa2025,Malizia2025,malizia2025disentangling}.

\textit{Discussion.} In this paper, we characterized in detail the impact of adaptive behaviors based on local awareness, driven either by pairwise or group interactions, on pairwise and higher-order contagion processes.
The adaptive mechanisms are based on the same principle in all cases: infectious nodes trigger awareness of risk in their neighborhoods, locally reducing the parameters determining contagion events and thus
the transmission rates. We have analyzed their effect through an analytical mean-field approach and numerical simulations, applied on a series of
empirical and synthetic hypergraph substrates.
In the active phase of the epidemic, in which the disease-free equilibrium is unstable, adaptive behaviours reduce the epidemic prevalence for both
types of contagion, with awareness based on group information having a stronger effect. 
However, their impact on the epidemic transition is very different for pairwise and higher-order processes. 
The former present a continuous phase transition in the non-adaptive case, and adaptive strategies do not change it, preserving its continuous nature and the epidemic threshold. Indeed, the triggering of adaptive behaviors tends to vanish when approaching the transition from above, as the epidemic prevalence itself goes to $0$. The higher-order contagion process instead presents a discontinuous phase transition with a bistability regime, driven by non-linear group infection mechanisms. Because of the discontinuity, the epidemic prevalence is large enough close to the transition, so that awareness mechanisms can be triggered and have an impact on the transition. 
As a consequence, adaptive behaviors are able to make the high-prevalence state unstable for $r$ not large enough, contracting the bistability region. If the strategy is efficient enough, it can even completely suppress higher-order effects and make the discontinuous transition and the bistability region disappear.
As adaptivity reduces the prevalence indeed, it decreases the possibility of occurrence of non-linear contagion events, defusing the explosiveness of group contagions that drive the discontinuous transition. 

Our work allows for a deeper understanding of higher-order processes on hypergraphs in the presence of adaptive behaviors, and makes possible a systematic comparison of the performance of different adaptive strategies on hypergraphs.
In particular, 
we show that strategies resulting from group-based awareness are more effective than those based on pairwise information. We explore both performance and costs of strategies in detail in \cite{mancastroppa2025}, considering a wide set of strategies based on local information of different nature, which can be pairwise, group, or hybrid.
Finally, our results can be the starting point for studying the impact of adaptive behaviors on other higher-order processes such as opinion formation and voter models \cite{lambiotte2022,voter2020,Schawe2022}, 
or for investigating the role of heterogeneity in higher-order adaptive mechanisms adoption \cite{mancastroppa2020active,Mancastroppa2021CT,mancastroppa2024}.

\textit{Data availability.} The data that support the findings of this study are publicly available. The SocioPatterns data sets are available at \cite{sociopatterns, conf_data}.

\textit{Acknowledgments.} M.M., M.K. and A.B. acknowledge support from the Agence Nationale de la Recherche (ANR) project DATAREDUX (ANR-19-CE46-0008). M.K. received support from the National Laboratory for Health Security (RRF-2.3.1-21-2022-00006); the MOMA WWTF project; and the BEQUAL NKFI-ADVANCED grant (153172).

\bibliographystyle{naturemag}
\bibliography{references_HO_adaptive_epi}

%\clearpage
\onecolumngrid
\begin{center}
    \section*{Appendix}
\end{center}
\twocolumngrid

\textit{Empirical and synthetic hypergraphs.} We consider several empirical and synthetic hypergraphs $\mathcal{H}$, and their corresponding weighted projected graphs 
$\mathcal{G}$. Projected graphs were obtained by projecting each hyperedge onto the corresponding clique and assigning each link a weight $w_l$ equal to the number of distinct hyperedges in which the link $l$ appears. 

We built static hypergraphs from empirical datasets describing face-to-face interactions in diverse environments: a hospital \cite{Vanhems2013,sociopatterns}, conferences \cite{Genois2023} and schools \cite{Stehle2011,sociopatterns,Mastrandrea2015} (see SM for details on each dataset and preprocessing). These hypergraphs cover a wide spectrum of properties (see SM), ranging from 76 to 327 nodes and from 315 to 4795 hyperedges of sizes up to 7. For example, the hospital dataset (considered in the main text) 
features $76$ nodes and $1102$ hyperedges of size $m \in [2,7]$.

Synthetic hypergraphs are obtained both by (i) applying randomization procedures to the empirical ones, with different reshuffling methods \cite{agostinelli2025higher,Mancastroppa2023} and (ii) setting specific features and mechanisms for hyperedges formation (such as specific hyper-degree distributions or overlap level). We refer to the SM for more details.

\textit{Numerical simulations and integration of mean-field equations.} We performed stochastic numerical simulations of the epidemic processes in discrete-time: infected individuals recover with probability $\mu$ per unit of time; a susceptible individual $i$ becomes infectious at time $t$ with probability $p_i(t)$. In the pairwise contagion process:
\begin{equation}
    p_i^p(t)=1-\prod\limits_{j \in \mathcal{V}} (1-\lambda_i(t) \lambda_j(t))^{w_{i,j}},
\end{equation}
which takes into account all possible contagion events and the multiplicity $w_l$ of the corresponding links. 
In the higher-order contagion process:
\begin{equation}
    p_i^{ho}(t)=1-\prod\limits_{e \in \mathcal{E}(i)} (1- \beta_e^i(i_e,\nu)),
\end{equation}
which takes into account all possible contagion events. In both processes, the simulations stop when the system reaches the endemic steady state or the absorbing state. If the system does not reach the absorbing state, the simulation is run for a minimum of $T$ time-steps and then the system is assumed to have reached the stationary state at the time-step $t>T$ if: (i) $\sigma(\bm{I}_{0.8t,t})<10^{-2}$ and (ii) $|\overline{\bm{I}_{0.9t,t}}-\overline{\bm{I}_{0.8t,0.9t}}|<10^{-2}$, where $\bm{I}_{a,b}=[|\mathcal{I}|(a),|\mathcal{I}|(a+1),...,|\mathcal{I}|(b-1),|\mathcal{I}|(b)]$ is the epidemic prevalence from time $a$ to $b$, $\sigma(\bm{I}_{a,b})$ is its standard deviation and $\overline{\bm{I}_{a,b}}$ is its average. The asymptotic epidemic prevalence is then calculated as $I_{\infty}=\overline{\bm{I}_{0.8t,t}}$.

The pairwise contagion process is simulated starting from a single infectious individual (selected uniformly at random) in a completely susceptible population. We set 
$\mu=10^{-2}$ and $T=1000$ without loss of generality and average the results over different initial conditions and over all outcomes of the epidemic (as long as at least one transmission event has occurred during the simulation). 
The higher-order process is simulated starting with either 2\% or 80\% of the population in the infectious state (infectious nodes are selected uniformly at random), in order 
to explore both branches of the phase diagram in the bistability region. In this case we use $\mu=10^{-1}$, $T=300$, and the results are averaged over different initial conditions and only over the realizations where the epidemic reaches an endemic state. For both type of processes, the results are averaged over 300 numerical simulations.

The mean-field equations are integrated numerically, considering the same epidemiological parameters as the numerical simulations. The pairwise process is integrated starting from $P_i(0)=0.01 \, \forall i$, while the higher-order
process is integrated either from $P_i(0)=0.01 \, \forall i$ or from 
$P_i(0)=0.8 \, \forall i$, to explore the two branches of the phase diagram.

\textit{Individual-based mean-field approach.} 
We show here that we obtain two closed sets of mean-field equations for the two contagion processes, by expressing all quantities as functions of the $P_i(t)$.
We first need to evaluate the probability $Q_{e \setminus i}(i_e,t)$ that in the hyperedge $e$ at time $t$ there are exactly $i_e$ infected individuals (other than $i$): 
\begin{equation}
    Q_{e \setminus i}(i_e,t)= \sum\limits_{A \in \mathcal{C}_{e \setminus i}(i_e)} \prod\limits_{j \in A} P_j(t) \prod\limits_{k \in e \setminus \{i,A\}} (1-P_k(t)),
    \label{eq:Q}
\end{equation}
where $\mathcal{C}_{e \setminus i}(i_e)$ is the set of all possible sets of $i_e$ nodes in $e \setminus i$. 

Analogously, we need to evaluate the term $\overline{\langle \lambda_j(t)/\lambda_0 \rangle_{i_e}^\nu}$, which involves a first average over the $i_e$ infected individuals in $e$, 
and then a second average over all possible configurations of $i_e$ infected individuals in $e \setminus i$:
\begin{equation}
    \overline{\langle \lambda_j(t)/\lambda_0 \rangle_{i_e}^\nu}= 
    \sum\limits_{A \in \mathcal{C}_{e \setminus i}(i_e)} H_A(t) 
    \left(\frac{1}{i_e} \sum\limits_{j \in A} e^{- f_j(t)} \right)^\nu,
\end{equation}
where:
\begin{equation}
    H_A(t)=\frac{\prod\limits_{j \in A} P_j(t) \prod\limits_{k \in e \setminus \{i,A\}} (1-P_k(t))}{\sum\limits_{B \in \mathcal{C}_{e \setminus i}(i_e)} \prod\limits_{j \in B} P_j(t) \prod\limits_{k \in e \setminus \{i,B\}} (1-P_k(t))}.
\end{equation}

To integrate the equations when adaptive strategies are taken into account, we need to evaluate
the awareness function $f_i(t)$. For the $nn$ strategy:  
\begin{equation}
    f_i^{nn}(t) = \sum\limits_{n=0}^{|\mathcal{N}(i)|} n Q_{\mathcal{N}(i)}(n,t)/\langle k \rangle ,
\end{equation}
where $Q_{\mathcal{N}(i)}(n,t) $ is the probability that $i$ has exactly $n$ infected neighbors at time $t$, which is calculated analogously to Eq. \eqref{eq:Q}. For the $ng$ strategy: 
\begin{equation}
    f_i^{ng}(t) = \sum\limits_{g=0}^{D(i)} g Z_{\mathcal{E}(i)}(g,t)/\langle D \rangle ,
\end{equation}
where $D(i)$ is the hyperdegree of node $i$, $\mathcal{E}(i)$ is the set of hyperedges in which $i$ is involved and $Z_{\mathcal{E}(i)}(g,t)$ is the probability of having $g$ infectious hyperedges at time $t$ among the hyperedges in $\mathcal{E}(i)$. 
We can write:
\begin{equation}
    Z_{\mathcal{E}(i)}(g,t)=\sum\limits_{A \in \mathcal{C}_{\mathcal{E}(i)}(g)} \prod\limits_{e \in A} \xi_{e,i}(t) \prod\limits_{h \in \mathcal{E}(i) \setminus A} (1-\xi_{h,i}(t)),
\end{equation}
where $\mathcal{C}_{\mathcal{E}(i)}(g)$ is the set of all possible sets of $g$ hyperedges
in $\mathcal{E}(i)$ and $\xi_{e,i}(t)$ is the probability that the hyperedge $e$ is infectious
at time $t$ for node $i$. In particular:
\begin{equation}
    \xi_{e,i}(t)=\sum\limits_{i_e=\phi_e(\theta)}^{|e|-1} Q_{e \setminus i}(i_e,t),
\end{equation}
where $\phi_e(\theta)=\lfloor \theta (|e|-1) \rfloor +1$ is the minimum number of infectious individuals in $e$ (distinct from $i$) so that $i$ perceives $e$ as at-risk.

\textit{Stability of the absorbing state.} By applying a linear stability analysis to Eq. \eqref{eq:MF_p}, we obtain a mean-field estimation of the epidemic threshold for the pairwise contagion case \cite{satorras2015epidemic}. 
In the NAD case, Eq. \eqref{eq:MF_p} reduces to: 
\begin{equation}
    \partial_t P_i(t)= -\mu P_i(t) + \lambda_0^2 (1-P_i(t)) \sum\limits_{j \in \mathcal{V}}  w_{i,j} P_j(t).
    \label{eq:MF_p_NAD}
\end{equation}
Linearizing Eq. \eqref{eq:MF_p_NAD} around the absorbing state $\bm{P}=[0,0,...,0]$ leads to the Jacobian matrix with element $J_{i,j}=-\mu \delta_{i,j} + \lambda_0^2 w_{i,j}$. The absorbing state is thus stable if and only if the largest eigenvalue of $\{J_{i,j}\}$ is negative, i.e. if:
\begin{equation}
   r=\lambda_0^2/\mu<1/\Lambda_w \equiv r_C^{NAD,p},
   \label{eq:conditions}
\end{equation}
where $\Lambda_w$ is the largest eigenvalue of $\{w_{i,j}\}$.

The adaptive strategies we considered feature the same Jacobian matrix as the NAD case. Indeed, expanding around the absorbing state and keeping only the linear leading order
yields:
$Q_{\mathcal{N}(i)}(n,t) \sim \delta_{n,1} \sum_{j \in \mathcal{N}(i)} P_j(t)$
and
$Q_{e \setminus i}(i_e,t) \sim \delta_{i_e,1} \sum_{j \in e \setminus i} P_j(t)$,
hence:
\begin{align}
    f_i^{nn}(t) &\sim \sum\limits_{j \in \mathcal{N}(i)} P_j(t)/\langle k \rangle,\\
    f_i^{ng}(t) &\sim \sum\limits_{e \in \mathcal{E}(i) | \phi_e(\theta) \leq 1} \sum\limits_{j \in e \setminus i} P_j(t)/\langle D \rangle.
\end{align}
Therefore, for both strategies: 
\begin{equation}
    \lambda_i(t) = \lambda_0 e^{- f_i(t)} \sim \lambda_0 (1+ o(\bm{P})) \forall i, \forall t.
    \label{eq:svilup}
\end{equation}
Inserting in Eq. \eqref{eq:MF_p}, we recover the NAD case of Eq. \eqref{eq:MF_p_NAD}, which implies that the epidemic thresholds for the pairwise contagion in the adaptive cases are the same as for the NAD case.

Moreover, the conditions for the absorbing state to be stable in the higher-order contagion process are fully determined by the pairwise interactions only \cite{FerrazdeArruda2021}, and are the same of the pairwise process (Eq. \eqref{eq:conditions}) independently on the adaptive strategy. Indeed, substituting the expansions around the absorbing state in Eq. \eqref{eq:MF_HO}, we obtain:
\begin{equation}
    \partial_t P_i(t) \sim -\mu P_i(t) + (1-P_i(t)) \lambda_0^2 \sum\limits_{e \in \mathcal{E}(i)}\sum\limits_{j \in e \setminus i} P_j(t),
\end{equation}
which correspond exactly to Eq. \eqref{eq:MF_p_NAD}, since $\sum_{e \in \mathcal{E}(i)} \sum_{j \in e \setminus i}$ can be rewritten as $\sum_{j \in \mathcal{V}} w_{i,j}$.%, given that each term is summed for each $j$ a number of times equal to the number of hyperedges in which both $i$ and $j$ are present. 

\textit{Quantifying the presence of non-linear contagion.}
Finally, we define the probability $\chi_e^{i_e>1}$ that a group $e$, containing at least a susceptible node, contains more than $1$ infected individual in the asymptotic steady state:
\begin{equation}
    \chi_e^{i_e>1}= \sum\limits_{i_e=2}^{|e|-1} Q_e(i_e,t \to \infty),
\end{equation}
with $Q_e(i_e,t \to \infty)$ calculated analogously to Eq. \eqref{eq:Q}. Then we can define the average probability $\langle \chi_e^{i_e>1} \rangle_m$ for groups of size $m$ and we can estimate the fraction of potential infections driven by non-linear group phenomena:
\begin{equation}
    \rho = \sum\limits_{m=2}^M \Psi(m) \frac{\langle \chi_e^{i_e>1} \rangle_m}{\langle \chi_e^{i_e>1} \rangle_m+\langle \chi_e^{i_e=1} \rangle_m},
\end{equation}
where $\chi_e^{i_e=1}=Q_e(1,t \to \infty)$, $M=\max\limits_{e \in \mathcal{E}}\{|e|\}$ and $\Psi(m)$ is the hyperedge size distribution.

\textit{From higher-order transition to a pairwise one.} 
We can analytically show, through the mean-field equations, that the higher-order contagion dynamics becomes equivalent to the pairwise one near the critical point, when the probability of having more than one infected agent in a hyperedge becomes close to zero. Let's indeed assume that $\chi_e^{i_e>1} \sim 0$, i.e. that the probability of having $i_e$ infected nodes in a hyperedge $e$ with one susceptible node at time $t$ is zero for $i_e>1$: 
\begin{equation}
    Q_{e \setminus i}(i_e,t) \sim \delta_{i_e,1} \sum\limits_{j \in e \setminus i} P_j(t) \prod\limits_{k \in e \setminus \{i,j\}} (1-P_k(t)).
    \label{eq:Q_aroundP}
\end{equation}
In this case, the epidemic dynamics reduces to: 
\begin{widetext}
\begin{equation}
    \partial_t P_i(t)= -\mu P_i(t) + (1-P_i(t)) \sum\limits_{e \in \mathcal{E}(i)}\sum\limits_{j \in e \setminus i} P_j(t) 
   \lambda_0 \lambda_i(t) (\lambda_j(t)/\lambda_0)^\nu 
    \prod\limits_{k \in e \setminus \{i,j\}} (1-P_k(t)).
    \label{eq:MF_HO_approx1}
\end{equation}
\end{widetext}
We replace $\sum_{e \in \mathcal{E}(i)} \sum_{j \in e \setminus i}$ with $\sum_{j \in \mathcal{V}} w_{i,j}$ and linearize the infection term around the absorbing state, using Eq. \eqref{eq:svilup}:
\begin{equation}
    \lambda_0 \lambda_i(t) (\lambda_j(t)/\lambda_0)^\nu
    \prod\limits_{k \in e \setminus \{i,j\}} (1-P_k(t)) \sim \lambda_0^2 .
\end{equation}
Therefore, by substituting in Eq. \eqref{eq:MF_HO_approx1}, if $\chi_e^{i_e>1} \sim 0$, the  contagion dynamics near the critical point is governed by: 
\begin{equation}
    \partial_t P_i(t) \sim -\mu P_i(t) + (1-P_i(t)) \sum\limits_{j \in \mathcal{V}} w_{i,j} \lambda_0^2 P_j(t),
    \label{eq:MF_HO_approx2}
\end{equation}
which reproduces that of the NAD pairwise case (see Eq. \eqref{eq:MF_p}), implying the continuous nature of the phase transition and the epidemic threshold $r_C=1/\Lambda_w$.

\end{document}

\title{Supplementary Material for "Adaptive behaviors neutralize bistable explosive transitions in higher-order contagion"}

\author{Marco Mancastroppa}
%\email{marco.mancastroppa@cpt.univ-mrs.fr}
\affiliation{Aix Marseille Univ, Université de Toulon, CNRS, CPT, Turing Center for Living Systems, 13009 Marseille, France}
\author{Márton Karsai}
\affiliation{Department of Network and Data Science, Central European University, 1100 Vienna, Austria}
\affiliation{National Laboratory for Health Security, HUN-REN Rényi Institute of Mathematics, 1053 Budapest, Hungary}
\author{Alain Barrat}
\affiliation{Aix Marseille Univ, Université de Toulon, CNRS, CPT, Turing Center for Living Systems, 13009 Marseille, France}

\maketitle

%\SItext

%\bigskip
%\tableofcontents
%\bigskip

In this Supplementary Material we present additional information and results. 
First, we present in detail in Section \ref{sez:section1} the empirical datasets considered, their preprocessing procedure, and their basic properties. In Section \ref{sez:section2}, we report results on the epidemic phase transition in the presence of adaptive behaviors for several datasets (as we 
present results only for one dataset in the main text), 
and we also report the local impact of these behaviors when 
there are few infected individuals. In Section \ref{sez:section3}, we present the procedure used to generate synthetic hypergraphs, either by reshuffling empirical hypergraphs or by synthetic generation, and we
report the main properties of the generated hypergraphs. Finally, in Section \ref{sez:section4}, we present results on epidemic propagation on synthetic hypergraphs and discuss the effect of hyperdegree heterogeneity and hyperedge overlap on the critical point of the phase transition, for pairwise and higher-order contagion processes
and the adaptive strategies considered.

\section{Empirical datasets}
\label{sez:section1}
We consider empirical datasets describing face-to-face interactions in several settings: a hospital (LH10 - \cite{Vanhems2013,sociopatterns}), two schools (Thiers13, LyonSchool - \cite{Stehle2011,sociopatterns,Mastrandrea2015}) and three conferences (WS16, ECIR19, ICCSS17 - \cite{Genois2023}). The datasets consist in time-resolved binary interactions, hence we need to preprocess the data to obtain static hypergraphs. We used a standard preprocessing method \cite{Mancastroppa2023,iacopini2022group}. The aggregate hypergraph $\mathcal{H}$ is built as follows: (i) all pairwise interactions are aggregated on time windows of duration $\Delta t$; (ii) all maximal cliques (not contained in other cliques) are promoted to hyperedges; (iii) only interactions observed at least $\delta$ times over the entire time span are retained as hyperedges of $\mathcal{H}$. 

In Supplementary Table \ref{tab:table1}, we report for each dataset 
the setting in which the interactions were collected, the parameters $\Delta t$ and $\delta$ of the preprocessing procedure and the main properties of the obtained empirical hypergraphs $\mathcal{H}$, such as number of nodes $N$, number of hyperedges $E$, average hyperedge size $\langle m \rangle$, average hyperdegree $\langle D \rangle$, average degree $\langle k \rangle$ and average link weight $\langle w_l \rangle$ in the weighted projected graph $\mathcal{G}$, and the heterogeneity of their distributions.
 
In Supplementary Fig. \ref{fig:figure1} we show for each dataset the hyperedge size distribution $\Psi(m)$, the distribution $P(D/\langle D \rangle)$ of the hyperedgree in $\mathcal{H}$, the distribution $P(k/\langle k \rangle)$ of the degree in $\mathcal{G}$ and the distribution $P(w_l/\langle w_l \rangle)$ of the link weights in $\mathcal{G}$.

The empirical datasets considered cover a wide range of different settings and present different statistical properties (see Supplementary Table \ref{tab:table1} and Supplementary Fig. \ref{fig:figure1}).

\begin{table}[h!]
    \begin{tabular}{|c|c|c|c|c|c|c|c|c|c|c|c|c|c|}
    \hline
    \hline
          & Setting & $\Delta t$ & $\delta$ & N & E & $\langle m \rangle$ & $\langle D \rangle$ & $\langle D^2 \rangle/\langle D \rangle^2$ & $\langle k \rangle$ & $\langle k^2 \rangle/\langle k \rangle^2$ & $\langle w_l \rangle$ & $\langle w_l^2 \rangle/\langle w_l \rangle^2$ \\ \hline
         LH10 & Hospital & 15 min &  1  &  76  &  1102  &  3.4 &  50.0  &  2.0  &  30.4  &  1.3  &  4.6  &  3.4\\
         Thiers13 & High-school & 15 min &  1  &  327  &  4795  &  3.1 &  45.3  &  1.3  &  35.6  &  1.1  &  3.0  &  2.2\\
         LyonSchool & Primary school & 15 min &  3  &  242  &  1188  &  2.4 &  11.8  &  1.2  &  13.9  &  1.2  &  1.3  &  1.3\\
         WS16 & Conference & 5 min &  2  &  137  &  315  &  3.4 &  7.8  &  1.3  &  16.1  &  1.3  &  1.3  &  1.3\\
         ECIR19 & Conference & 5 min &  2  &  169  &  708  &  2.7  &  11.3  &  1.3  &  17.3  &  1.3  &  1.3  &  1.4 \\
         ICCSS17 & Conference & 5 min &  2  &  271  &  1473  &  2.6 &  13.9  &  1.3  &  20.4  &  1.4  &  1.2  &  1.3\\
    \hline
    \hline
    \end{tabular}
    \caption{\textbf{Properties of empirical datasets.} For each empirical dataset we indicate: the setting in which the face-to-face interactions have been collected, the preprocessing parameters $\Delta t$ and $\delta$, the number of nodes $N$, the number of hyperedges $E$, their average size $\langle m \rangle$, the average hyperdegree $\langle D \rangle$ and the heterogeneity $\langle D^2 \rangle/\langle D \rangle^2$ of its distribution, the average degree $\langle k \rangle$ and the heterogeneity $\langle k^2 \rangle/\langle k \rangle^2$ of its distribution, the average weight $\langle w_l \rangle$ in the projected graph $\mathcal{G}$ and the heterogeneity $\langle w_l^2 \rangle/\langle w_l \rangle^2$ of its distribution. Note that the statistics of the weights in $\mathcal{G}$ considers only existing links, so $\langle w_l \rangle \geq 1$.} 
    \label{tab:table1}
\end{table}

\begin{figure*}[ht!]
\includegraphics[width=0.97\textwidth]{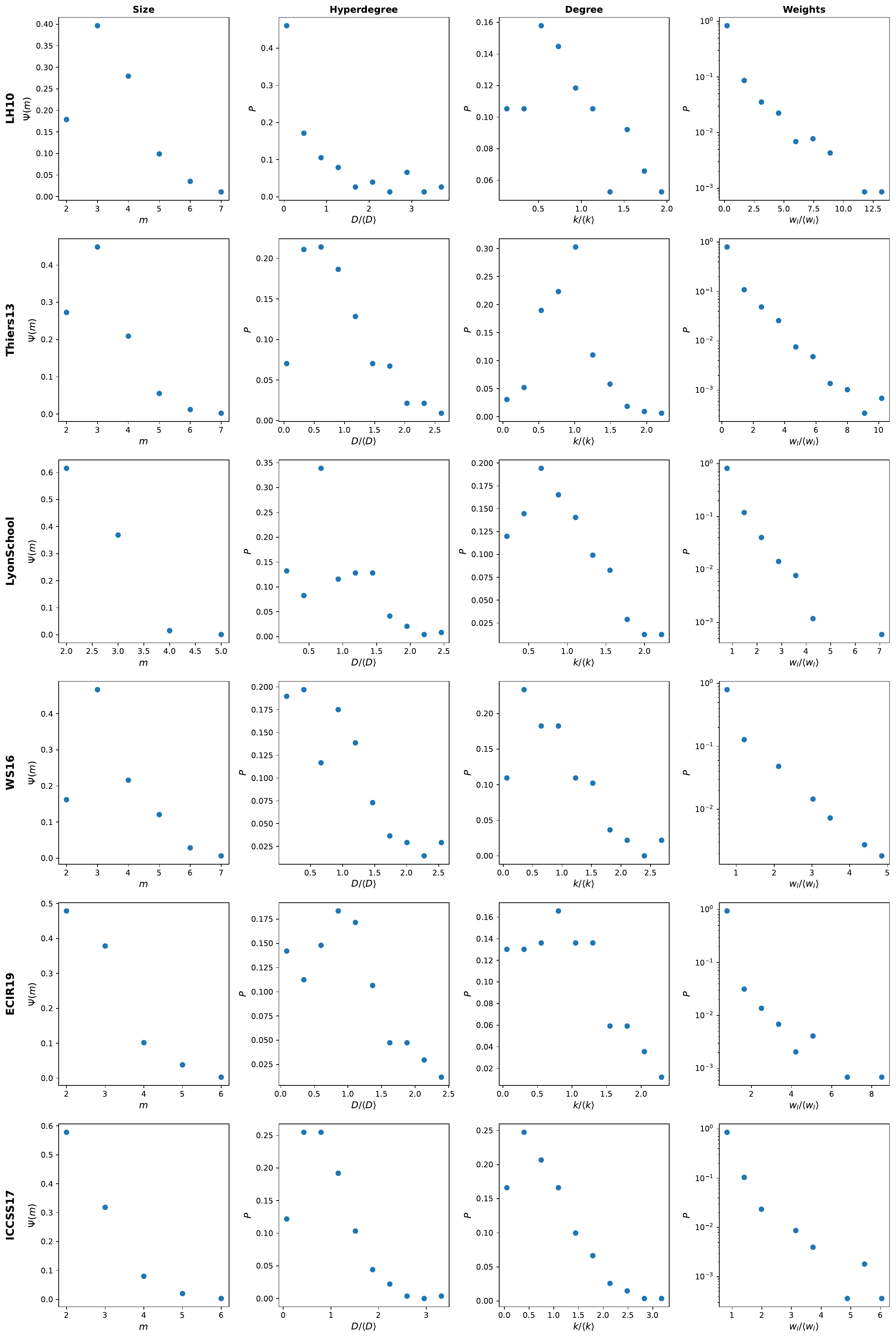}
\caption{\textbf{Properties of empirical datasets.} Each row corresponds to a different dataset (see labels): for each of them the first column shows the hyperedge size distribution $\Psi(m)$; the second column the distribution $P(D/\langle D \rangle)$ of the hyperdegree in $\mathcal{H}$; the third column the distribution $P(k/\langle k \rangle)$ of the degree in $\mathcal{G}$; the fourth column the distribution $P(w_l/\langle w_l \rangle)$ of the link weights in $\mathcal{G}$.}
\label{fig:figure1}
\end{figure*}

\clearpage
\section{Results on empirical datasets}
\label{sez:section2}
\begin{figure*}[ht!]
\includegraphics[width=0.9\textwidth]{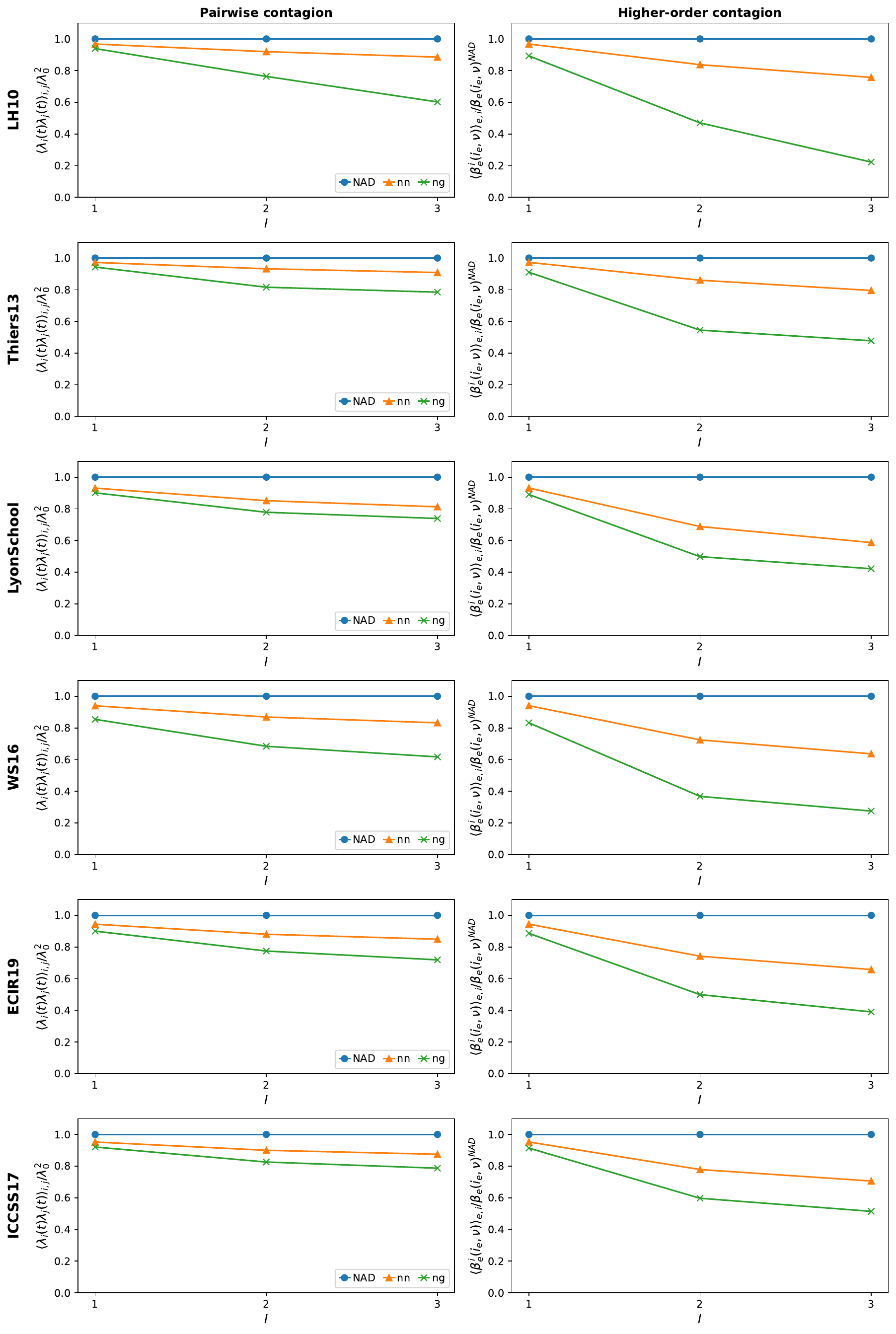}
\caption{\textbf{Local effects of adaptive behaviours - Empirical.} Each row corresponds to a different dataset (see labels): for each of them the first column focuses 
on the pairwise contagion and we show the average reduction in contagion probability $\langle \lambda_i(t) \lambda_j(t) \rangle_{i,j}/\lambda_0^2$ for links connecting an infected node to a healthy one for the different adaptive strategies. In the second column, we focus on higher-order contagion and show the average reduction in contagion probability $\langle \beta_e^i(i_e,\nu) \rangle_{e,i}/\beta_e(i_e,\nu)^{NAD}$ for a susceptible node within a hyperedge with at least one infected node for the different adaptive strategies. In both panels, we consider $I=[1, 2, 3]$ infected nodes, each point is averaged over all edges and all possible combinations of infected nodes, $\nu=4$, $\lambda_0=0.02$ and $\theta=0$ for the $ng$ strategy.
}
\label{fig:figure2}
\end{figure*}

Here we report results on the epidemic phase transition in the presence of adaptive behaviours for pairwise and higher-order contagion processes on the 
empirical datasets. 
In particular, Supplementary Fig. \ref{fig:figure2} shows the local impact of the adaptive behaviours when there are only few infected individuals: we estimate it through the local reduction of the infection probability for links (in the pairwise contagion) or groups (in the higher-order contagion). In Supplementary Fig. \ref{fig:figure3} we report the epidemic phase diagram for the different adaptive strategies, for the pairwise and higher-order contagion process on empirical datasets not shown in the main text. 
In Supplementary Fig. \ref{fig:figure4} we show how bistability and discontinuity are neutralized by adaptive behaviours, showing the probability $\langle \chi_e^{i_e>1} \rangle_m$ that a group of size $m$ with at least a susceptible node has more than $1$ infected individuals in the asymptotic steady state, as a function of the effective infection rate for different values of $m$, and the probability $\rho$ that a group of arbitrary size supports higher-order contagion. 

These results are analogous to those presented in the main text for the hospital dataset, even if obtained for empirical systems very different between each other, 
in terms of settings and statistical properties (see Section \ref{sez:section1}), showing the robustness of our findings.
Note that, for some datasets both $nn$ and $ng$ strategies completely suppresses
the bistability region, while in other cases both lead to a still discontinuous 
transitions with a shrinked but still finite
bistability.

\begin{figure*}[ht!]
\includegraphics[width=0.8\textwidth]{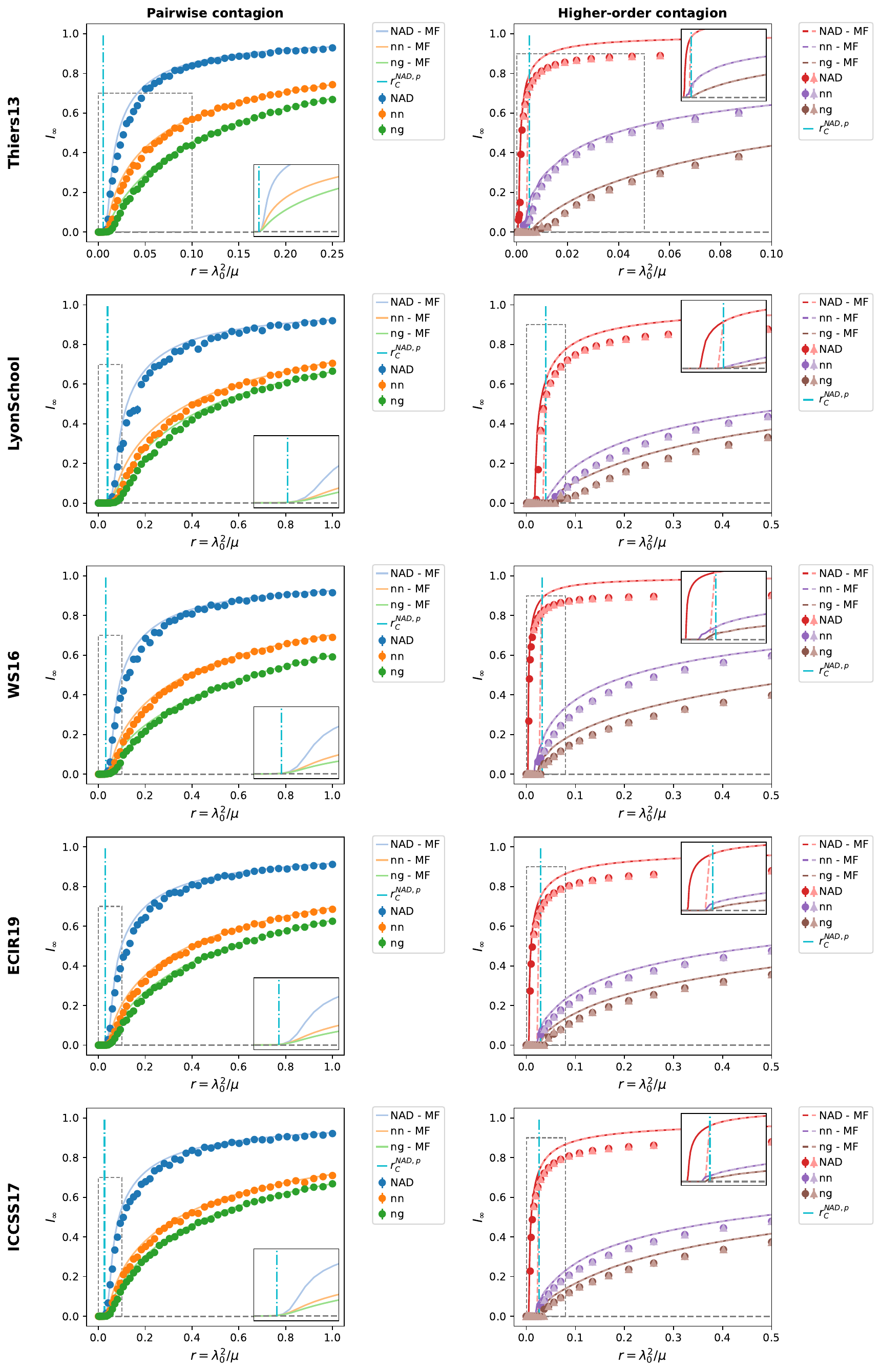}
\caption{\textbf{Phase diagram for the pairwise and higher-order contagion processes - Empirical.} Each row corresponds to a different dataset (see labels): for each of them we show the epidemic prevalence in the asymptotic steady state $I_{\infty}$ as a function of the effective infection rate $r$, for the non-adaptive case (NAD) and for the two adaptation strategies ($nn$, $ng$), respectively for the pairwise (first column) and higher-order contagion process (second column). For each case, we show the results of numerical simulations (markers) and of integration of the mean-field equations (lines). For the higher-order contagion case, we consider simulations and numerical integration starting with a low (triangles and dashed lines) and a high (dots and solid lines) fraction of infected nodes. In both panels, the simulation results are averaged over $200$ simulations, $\nu=4$, $\theta=0.3$, the inset represents a zoom on the mean-field results in the area marked by a dashed rectangle and the vertical dash-dotted line indicate $r_C^{NAD,p}$. For numerical simulations, error-bars correspond to one standard deviation (sometimes they are not visible since smaller than the corresponding markers).}
\label{fig:figure3}
\end{figure*}

\begin{figure*}[ht!]
\includegraphics[width=0.9\textwidth]{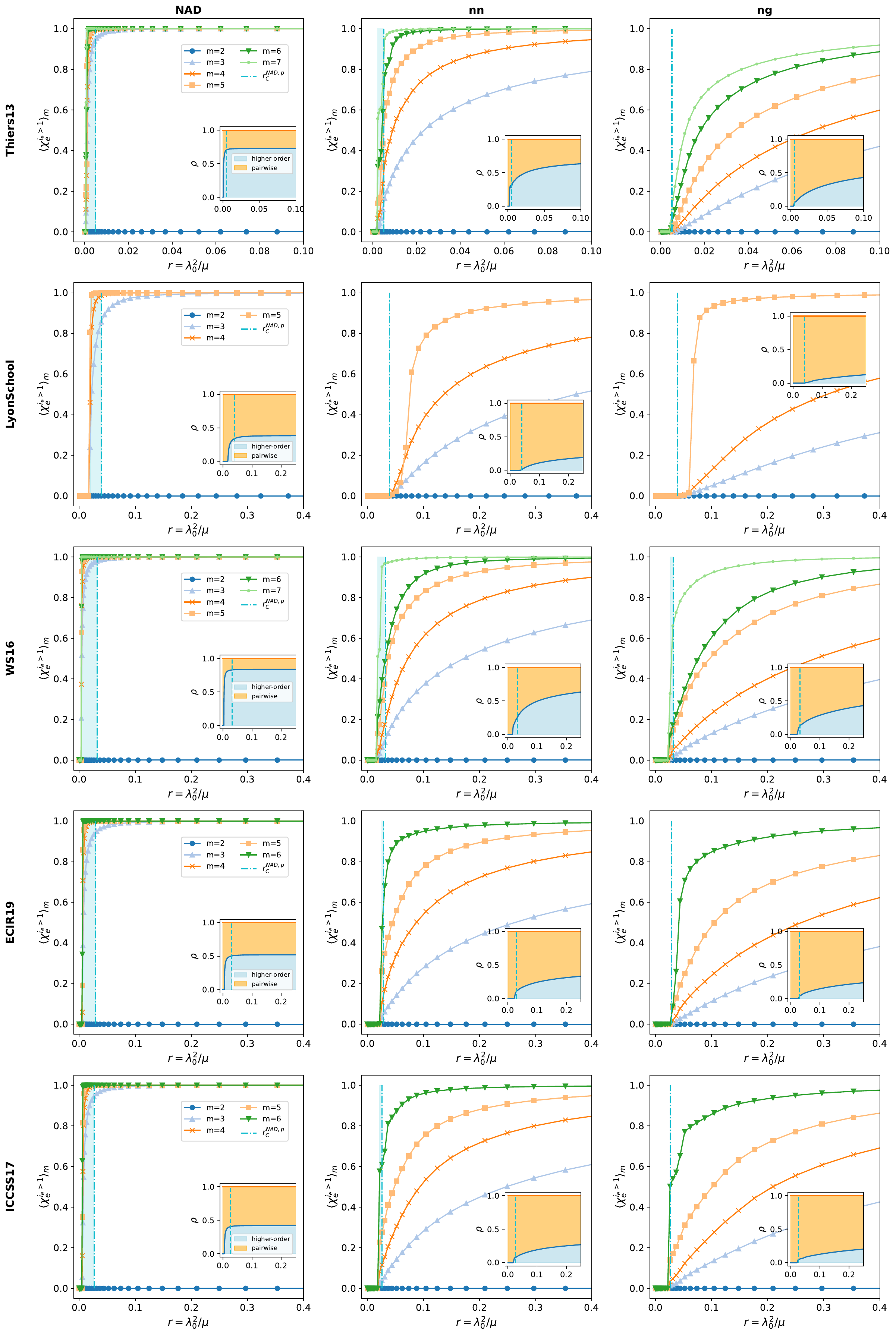}
\caption{\textbf{Neutralization of higher-order contagion - Empirical.} Each row corresponds to a different dataset (see labels): for each of them in each column, we show the probability $\langle \chi_e^{i_e>1} \rangle_m$ that a group of size $m$ with at least one susceptible node has more than $1$ infected individuals in the asymptotic steady state, as a function of the effective infection rate for different values of $m$. We estimate it through numerical integration of the mean-field equations starting from a high fraction of infected nodes. Each column corresponds to an adaptive strategy (with $\nu=4$ and $\theta=0.3$ for the $ng$ strategy), the vertical dash-dotted line indicates $r_C^{NAD,p}$ and the light-blue colored area indicates the bistability region. In the inset, we show the probability $\rho$ that a group of arbitrary size supports higher-order contagion (light-blue area) as a function of $r$, and the complementary probability that it supports pairwise contagion (orange area).}
\label{fig:figure4}
\end{figure*}

\clearpage
\section{Synthetic hypergraphs}
\label{sez:section3}
We consider three ways of building synthetic hypergraphs, obtained starting from an empirical one 
$\mathcal{H}$:
\begin{itemize}
    \item \textbf{G}: we construct a hypergraph $\mathcal{H}'$ by projecting each hyperedge $e \in \mathcal{H}$ onto its clique and without considering link weights. This procedure generates an unweighted pairwise graph, with the same number of nodes $N$ and same degree distribution $P(k)$ as the original hypergraph, but destroys any higher-order properties of $\mathcal{H}$, including the hyperedge size distribution $\Psi(m)$;
    \item \textbf{ER}: we build an Erdős–Rényi-like randomization of the empirical hypergraph, $\mathcal{H}'$ \cite{agostinelli2025higher}. The generated hypergraph has the same number $N$ of nodes and $E_m$ of hyperedges of size $m$ 
    as the empirical hypergraph $\mathcal{H}$: each hyperedge in $\mathcal{H}'$ is created by selecting its nodes uniformly at random from the population. This randomization preserves the number of nodes and the number of hyperedges of each size (and therefore also the hyperedges size distribution $\Psi(m)$), but destroys the distribution of hyperdegrees $P(D)$ (overall and for each size) and any macro- and meso-structure (e.g. communities, hypercores).
    \item \textbf{k}: we construct a randomization of the empirical hypergraph, $\mathcal{H}'$, in the spirit of the configuration model for graphs \cite{Mancastroppa2023,agostinelli2025higher}. Specifically, $\mathcal{H}'$ has the same number of nodes $N$ and hyperedges $E_m$ for each size $m$ as the original hypergraph $\mathcal{H}$: each hyperedge in $\mathcal{H}$ is created by selecting its nodes with probability proportional to their hyperdegree $D$ in the original hypergraph $\mathcal{H}$. This randomization preserves the number of nodes, the number of hyperedges of each size (and therefore also the size distribution $\Psi(m)$), and the hyperdegree distribution $P(D)$, but randomizes how different nodes participate in the different sizes (hyperdegrees by order) and any macro- and meso-structure (e.g. communities, hypercores).
\end{itemize}

We also consider two types of synthetic hypergraphs obtained by imposing specific properties, instead of randomizing interactions in the empirical datasets. We consider a hypergraph $\mathcal{H}'$ with $N$ nodes and $E$ hyperedges, whose sizes are 
drawn from the distribution of hyperedge sizes $\Psi(m)$:
\begin{itemize}
    \item \textbf{PD}: this model generates a hypergraph $\mathcal{H}'$ with a specific hyperdegree distribution $P(D)$. Given the hyperdegree distribution $P(D)$ desired, we extract a parameter $x$ for each node from a distribution $P(x)$ with the same functional form as $P(D)$; then we generate the hyperedges of $\mathcal{H}'$ by selecting nodes with a probability proportional to $x$. In particular, we consider three cases: a Dirac Delta $P(x) \sim \delta(x-1)$, for which all nodes have the same probability of participating in hyperedges; an exponential $P(x) \sim e^{-x/\beta}$ for various values of $\beta$ (see Supplementary Table \ref{tab:table2}); a power-law $P(x) \sim x^{\alpha-1}$ (with lower cut-off at $10^{-3}$ and upper cut-off at $1$) for various values of $\alpha$ (see Supplementary Table \ref{tab:table2}). The hypergraph $\mathcal{H}'$ will have a hyperdegree distribution $P(D)$ with approximately the same shape as $P(x)$, whose average, minimum and maximum values depend on the number of hyperedges $E$ and of nodes $N$ chosen.
    \item \textbf{OV}: generates a hypergraph $\mathcal{H}'$ with a tunable level of hyperedges overlap, measured as $\langle w_l \rangle$. We initially divide the population randomly into $n$ communities of approximately the same size. Each hyperedge $e$ of size $m$ (drawn from $\Psi(m)$) in $\mathcal{H}'$ is created as follows: (i) the first node in $e$ is selected uniformly at random in the whole population; (ii) the other $m-1$ nodes are selected with a probability proportional to $p$ if they belong to the same community, and with a probability proportional to $1-p$ if they belong to a different community. The generated hypergraph $\mathcal{H}'$ has a homogeneous distribution of the hyperdegree $P(D)$ and the value of $p$ tunes the overlap level: $p=0.5$ corresponds to a Erdős–Rényi-like hypergraph, since the nodes participating in an hyperedge are selected with the same probability regardless of their community; by increasing $p$ the overlap level is progressively increased, since interactions within the same community are favoured; $p \sim 1$ generates highly clustered hypergraphs with high overlap level.
\end{itemize}
For simplicity and to allow comparison with the others synthetic hypergraphs, we fix the number of nodes $N$ and the number of hyperedges $E_m$ of each size $m$ equal to those of an empirical hypergraph $\mathcal{H}$, thus reproducing its size distribution $\Psi(m)$.

We generate synthetic hypergraphs with these five methods starting from the hospital dataset (LH10), whose results have been shown in the main text. In the Supplementary Table \ref{tab:table2}, we show the main statistical properties of the synthetic hypergraphs generated, specifying the generation parameters where necessary. Moreover, in Supplementary Fig. \ref{fig:figure5} we report for selected synthetic hypergraphs: the hyperedge size distribution $\Psi(m)$, the distribution $P(D/\langle D \rangle)$ of the hyperedgree in $\mathcal{H}$, the distribution $P(k/\langle k \rangle)$ of the degree in $\mathcal{G}$ and the distribution $P(w_l/\langle w_l \rangle)$ of the link weights in $\mathcal{G}$.

\begin{table}[h!]
    \begin{tabular}{|c|c|c|c|c|c|c|c|}
    \hline
    \hline
           & $\langle m \rangle$ & $\langle D \rangle$ & $\langle D^2 \rangle/\langle D \rangle^2$ & $\langle k \rangle$ & $\langle k^2 \rangle/\langle k \rangle^2$ & $\langle w_l \rangle$ & $\langle w_l^2 \rangle/\langle w_l \rangle^2$ \\ \hline
         Data - LH10 & 3.4 & 50.0 & 2.0 & 30.4 & 1.3 & 4.6 & 3.4\\
         G & 2.0 & 30.4 & 1.3 & 30.4 & 1.3 & 1.0 & 1.0 \\
         ER & 3.4 & 50.0 & 1.0 & 62.9 & 1.0 & 2.2 & 1.3 \\
         k & 3.4 & 50.0 & 2.0 & 39.7 & 1.2 & 3.5 & 2.4 \\
         PD - Dirac delta & 3.4 & 50.0 & 1.0 & 63.4 & 1.0 & 2.2 & 1.3\\
         PD - Exp $\beta=0.5$ & 3.4 & 52.8 & 1.8 & 39.6 & 1.2 & 3.7 & 2.1 \\
         PD - Exp $\beta=1.5$ & 3.4 & 52.8 & 1.7 & 40.2 & 1.2 & 3.6 & 1.9 \\
         PD - Exp $\beta=3.5$ & 3.4 & 52.1 & 2.0 & 36.8 & 1.2 & 3.9 & 2.3\\
         PD - Power-law $\alpha=0.01$ & 3.4 & 52.8 & 2.8 & 27.4 & 1.4 & 5.4 & 3.3 \\
         PD - Power-law $\alpha=0.75$ & 3.4 & 53.5 & 1.4 & 43.3 & 1.2 & 3.4 & 1.6 \\
         PD - Power-law $\alpha=1.5$ & 3.4 & 52.0 & 1.2 & 54.5 & 1.0 & 2.7 & 1.4 \\
         PD - Power-law $\alpha=10$ & 3.4 & 52.0 & 1.0 & 61.7 & 1.0 & 2.3 & 1.3 \\
         OV - $n=7$, $p=0.5$ & 3.4  &  50.0  &  1.0  &  62.9  &  1.0  &  2.2  &  1.3 \\
         OV - $n=7$, $p=0.95$ & 3.4  &  50.0  &  1.0  &  47.4  &  1.0  &  2.9  &  2.2\\
         OV - $n=7$, $p=0.98$ & 3.4  &  50.0  &  1.0  &  34.1  &  1.0 &  4.1  &  2.4\\
         OV - $n=7$, $p=0.99$ & 3.4 & 50.0 & 1.0 & 21.6 & 1.0 & 6.4 & 1.9\\
    \hline
    \hline
    \end{tabular}
    \caption{\textbf{Properties of synthetic hypergraphs.} For each synthetic hypergraph generated we indicate: the parameters for its generation (if needed), the average hyperedge size $\langle m \rangle$, the average hyperdegree $\langle D \rangle$ and the heterogeneity of its distribution $\langle D^2 \rangle/\langle D \rangle^2$, the average degree $\langle k \rangle$ and the heterogeneity of its distribution $\langle k^2 \rangle/\langle k \rangle^2$, and the average weight $\langle w_l \rangle$ in the projected graph $\mathcal{G}$ and the heterogeneity of its distribution $\langle w_l^2 \rangle/\langle w_l \rangle^2$. Note that the statistics of the weights in $\mathcal{G}$ consider only existing links, so $\langle w_l \rangle \geq 1$. Moreover, for all the hypergraphs the numbers of nodes $N$ and of hyperedges $E_m$, for each size $m$, are as in the empirical LH10 dataset (see Supplementary Table \ref{tab:table1}).} 
    \label{tab:table2}
\end{table}

\begin{figure*}[ht!]
\includegraphics[width=\textwidth]{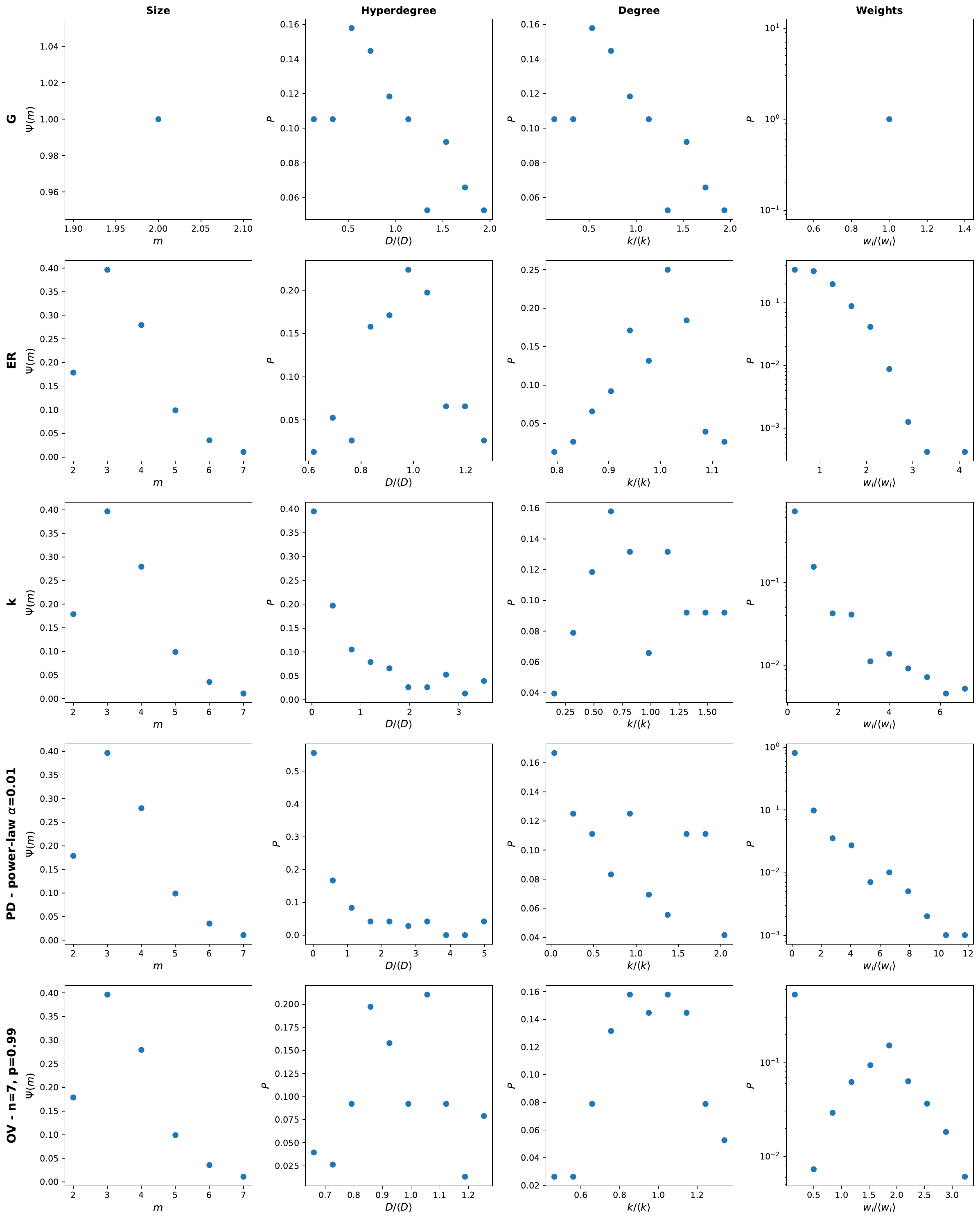}
\caption{\textbf{Properties of synthetic hypergraphs.} Each row corresponds to a different synthetic hypergraph (see labels): for each of them the first column show the hyperedge size distribution $\Psi(m)$; the second column the distribution $P(D/\langle D \rangle)$ of the hyperedgree in $\mathcal{H}$; the third column the distribution $P(k/\langle k \rangle)$ of the degree in $\mathcal{G}$; the fourth column the distribution $P(w_l/\langle w_l \rangle)$ of the link weights in $\mathcal{G}$ (considering only non-zero $w_l$).}
\label{fig:figure5}
\end{figure*}

\clearpage
\section{Results on synthetic hypergraphs}
\label{sez:section4}
Here we report results regarding the epidemic phase transition in the presence of adaptive behaviours for pairwise and higher-order contagion processes on the generated synthetic hypergraphs. 

\begin{figure*}[ht!]
\includegraphics[width=\textwidth]{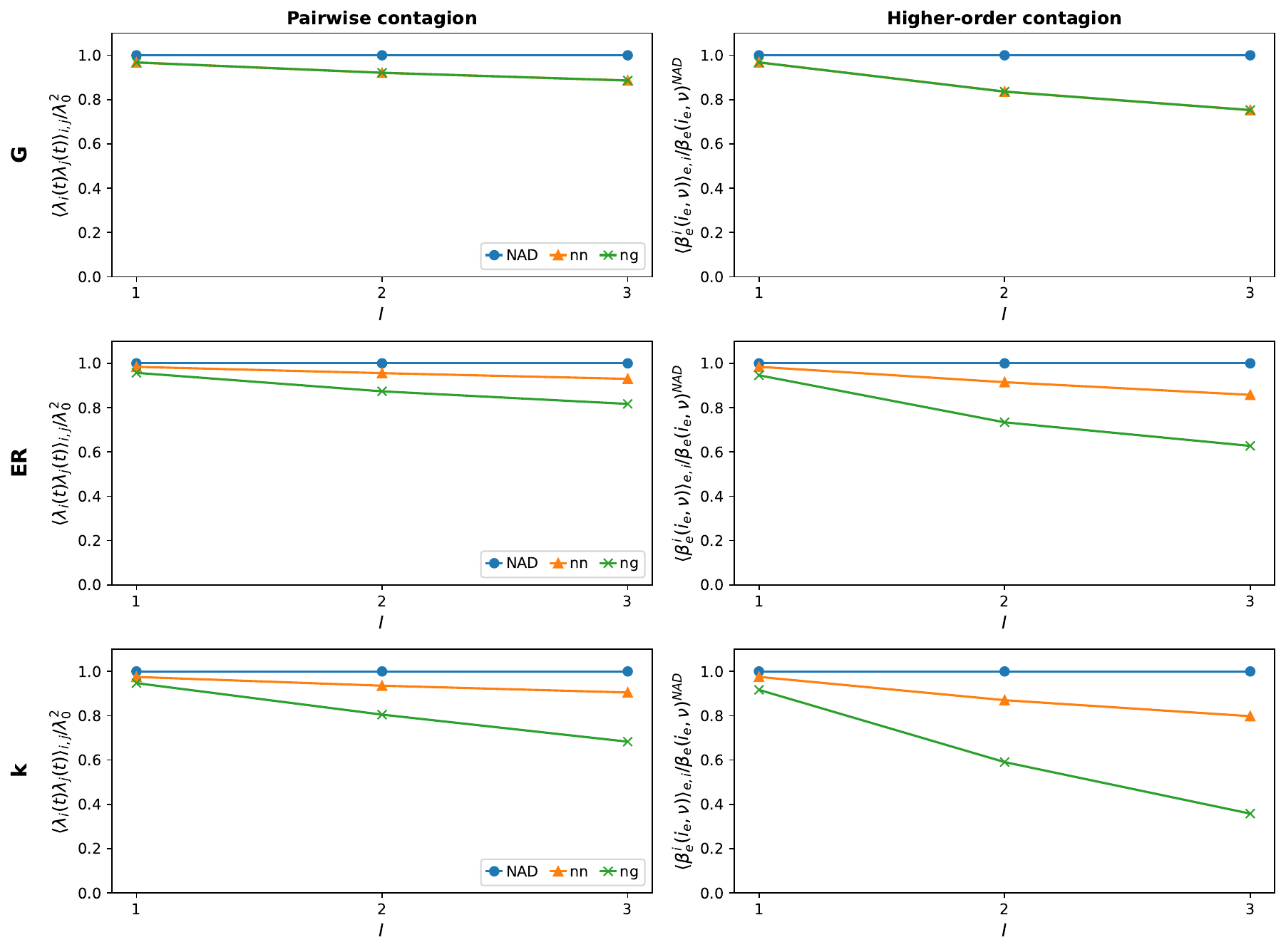}
\caption{\textbf{Local effects of adaptive behaviours - Randomized and projected.} Each row corresponds to a different randomization or projection of the hypergraph obtained from the hospital dataset (see labels): for each of them the first column focus on the pairwise contagion and we show the average reduction in contagion probability $\langle \lambda_i(t) \lambda_j(t) \rangle_{i,j}/\lambda_0^2$ for links connecting an infected node to a healthy one for the different adaptive strategies. In the second column, we focus on higher-order contagion and show the average reduction in contagion probability $\langle \beta_e^i(i_e,\nu) \rangle_{e,i}/\beta_e(i_e,\nu)^{NAD}$ for a susceptible node within a hyperedge with at least one infected node for the different adaptive strategies. In both panels, we consider $I=[1, 2, 3]$ infected nodes, each point is averaged over all edges and all possible combinations of infected nodes, $\nu=4$, $\lambda_0=0.02$ and $\theta=0$ for the $ng$ strategy.
}
\label{fig:figure6}
\end{figure*}

\begin{figure*}[ht!]
\includegraphics[width=\textwidth]{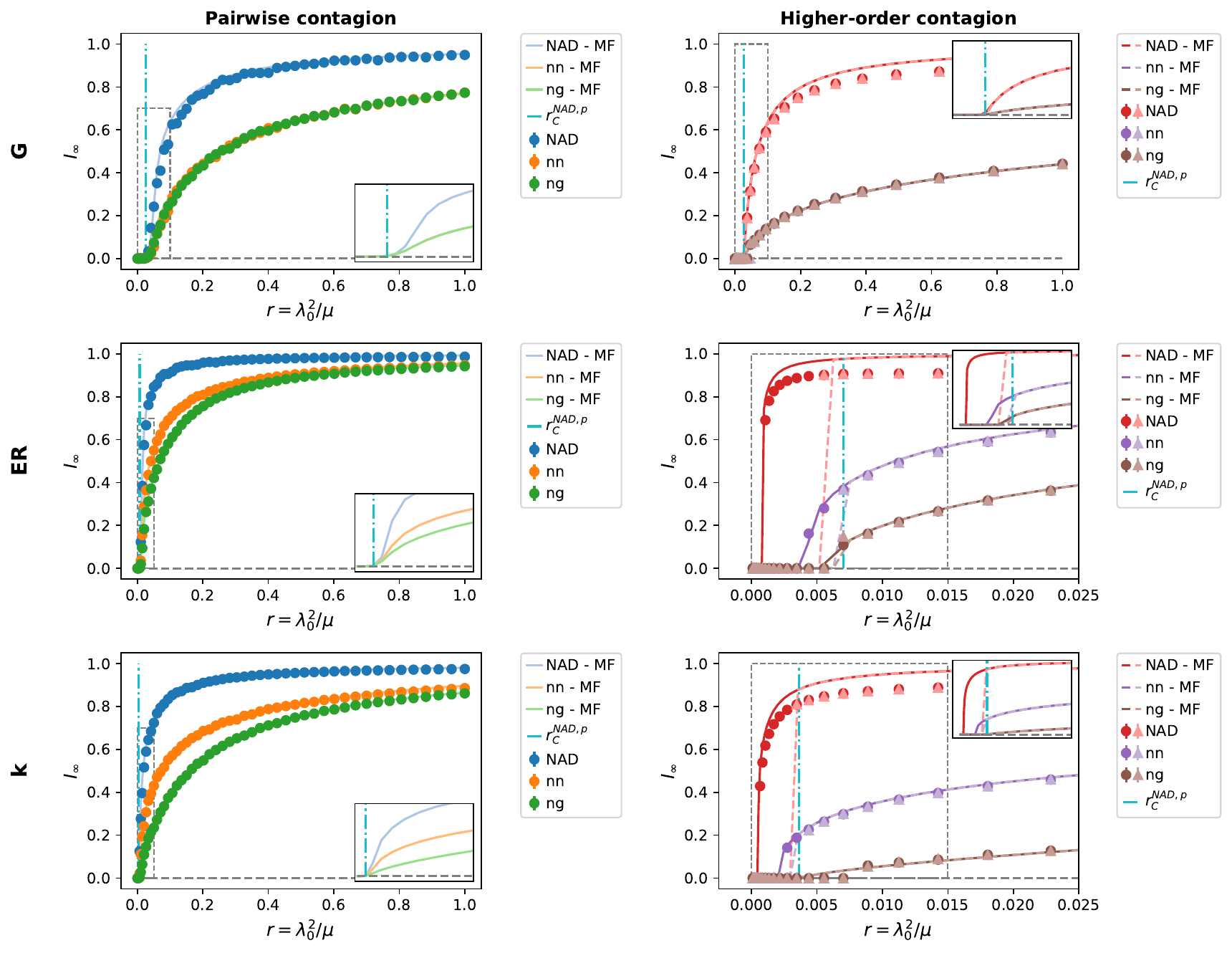}
\caption{\textbf{Phase diagram for the pairwise and higher-order contagion processes - Randomized and projected.} Each row corresponds to a different randomization or projection of the hypergraph obtained from the hospital dataset (see labels): for each of them we show the epidemic prevalence in the asymptotic steady state $I_{\infty}$ as a function of the effective infection rate $r$, for the non-adaptive case (NAD) and for the two adaptation strategies ($nn$, $ng$), respectively for the pairwise (first column) and higher-order contagion process (second column). For each case, we show the results of numerical simulations (markers) and of integration of the mean-field equations (lines). For the higher-order contagion case, we consider simulations and numerical integration starting with a low (triangles and dashed lines) and a high (dots and solid lines) fraction of infected nodes. In both panels, the simulation results are averaged over $200$ simulations, $\nu=4$, $\theta=0.3$, the inset represents a zoom on the mean-field results in the area marked by a dashed rectangle and the vertical dash-dotted line indicate $r_C^{NAD,p}$. For numerical simulations, error-bars correspond to one standard deviation (sometimes they are not visible since smaller than the corresponding markers).}
\label{fig:figure7}
\end{figure*}

\begin{figure*}[ht!]
\includegraphics[width=\textwidth]{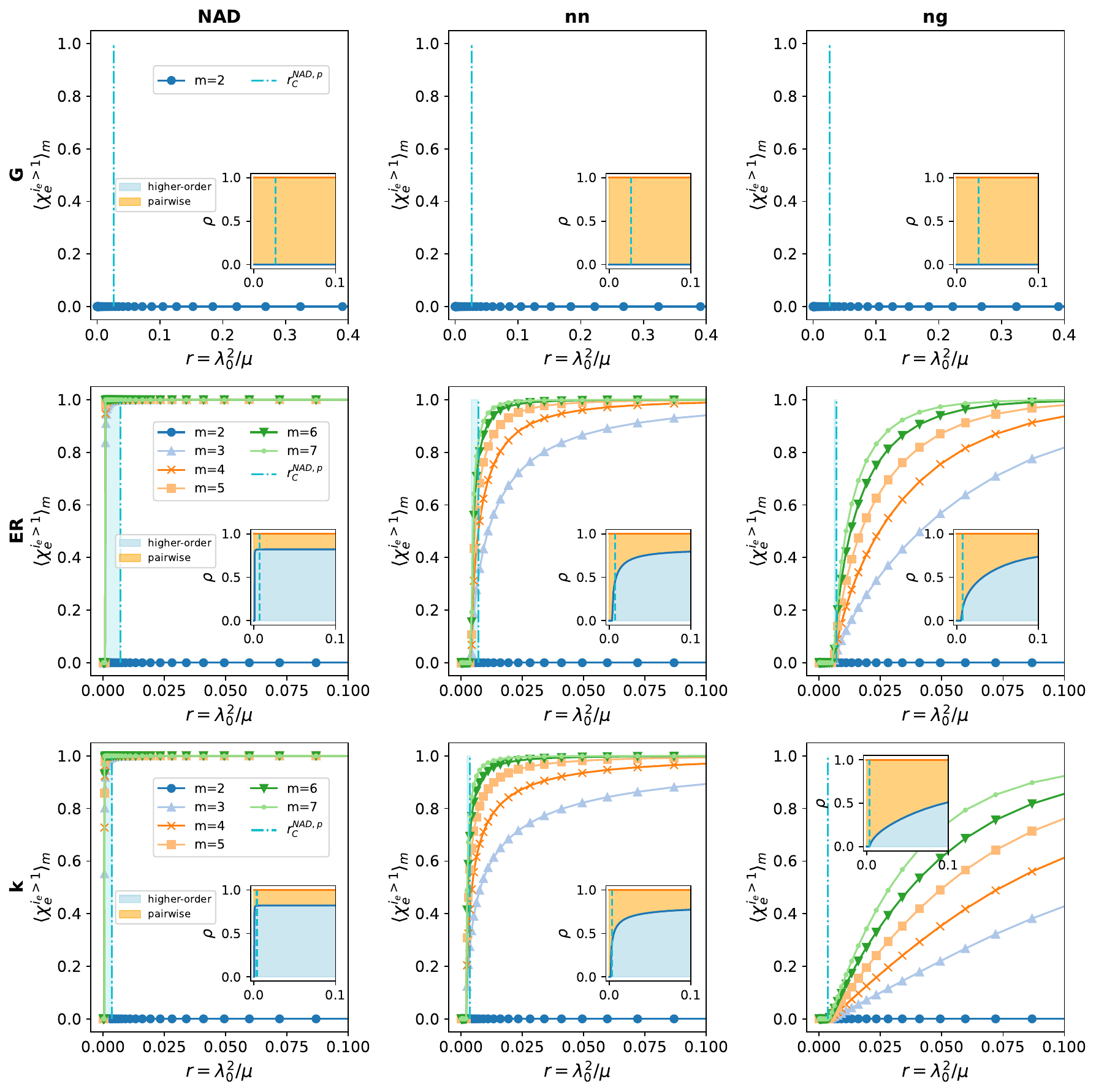}
\caption{\textbf{Neutralization of higher-order contagion - Randomized and projected.} Each row corresponds to a different randomization or projection of the hypergraph obtained from the hospital dataset (see labels): for each of them in each column, we show the probability $\langle \chi_e^{i_e>1} \rangle_m$ that a group of size $m$ with at least one susceptible node has more than $1$ infected individuals in the asymptotic steady state, as a function of the effective infection rate for different values of $m$. We estimate it through numerical integration of the mean-field equations starting from a high fraction of infected nodes. Each column corresponds to an adaptive strategy (with $\nu=4$ and $\theta=0.3$ for the $ng$ strategy), the vertical dash-dotted line indicates $r_C^{NAD,p}$ and the light-blue colored area indicates the bistability region. In the inset, we show the probability $\rho$ that a group of arbitrary size supports higher-order contagion (light-blue area) as a function of $r$, and the complementary probability that it supports pairwise contagion (orange area).}
\label{fig:figure8}
\end{figure*}

Supplementary Fig. \ref{fig:figure6} shows the local impact of the adaptive behaviours when there are a few infected individuals in the randomized and projected hypergraphs. Supplementary Fig.s \ref{fig:figure7}, \ref{fig:figure8} show respectively the epidemic phase diagram for the different adaptive strategies, for the pairwise and higher-order contagion processes on the randomized and projected hypergraphs, and the probability $\langle \chi_e^{i_e>1} \rangle_m$ as a function of the effective infection rate for different values of $m$, showing how bistability and discontinuity are neutralized by adaptive behaviours. These results allow us to better understand the effect of the underlying higher-order structure on contagion processes. If the underlying structure is a simple graph (G), the adaptive strategies $nn$ and $ng$ coincide and are equivalent, both near the critical point and in the active phase of the epidemic; furthermore, the higher-order contagion process becomes pairwise, since there are no groups. On the contrary, the ER-hypergraph and the k-hypergraph present dynamics qualitatively similar to the empirical case for both the pairwise and higher-order processes: 
the underlying structure, however, quantitatively influences the phase transition and the active phase. For example, the ER case has a wider bistability region than the k case, which instead is more similar to the empirical case, as it preserves 
more of its structural properties. 
Furthermore, in the ER case, the $nn$ and $ng$ strategies are more similar to each other than in the empirical case and in the k case. Indeed, what makes the two strategies different is how they exploit the heterogeneity of the population, as we investigate in \cite{mancastroppa2025}.\\

\begin{figure*}[ht!]
\includegraphics[width=\textwidth]{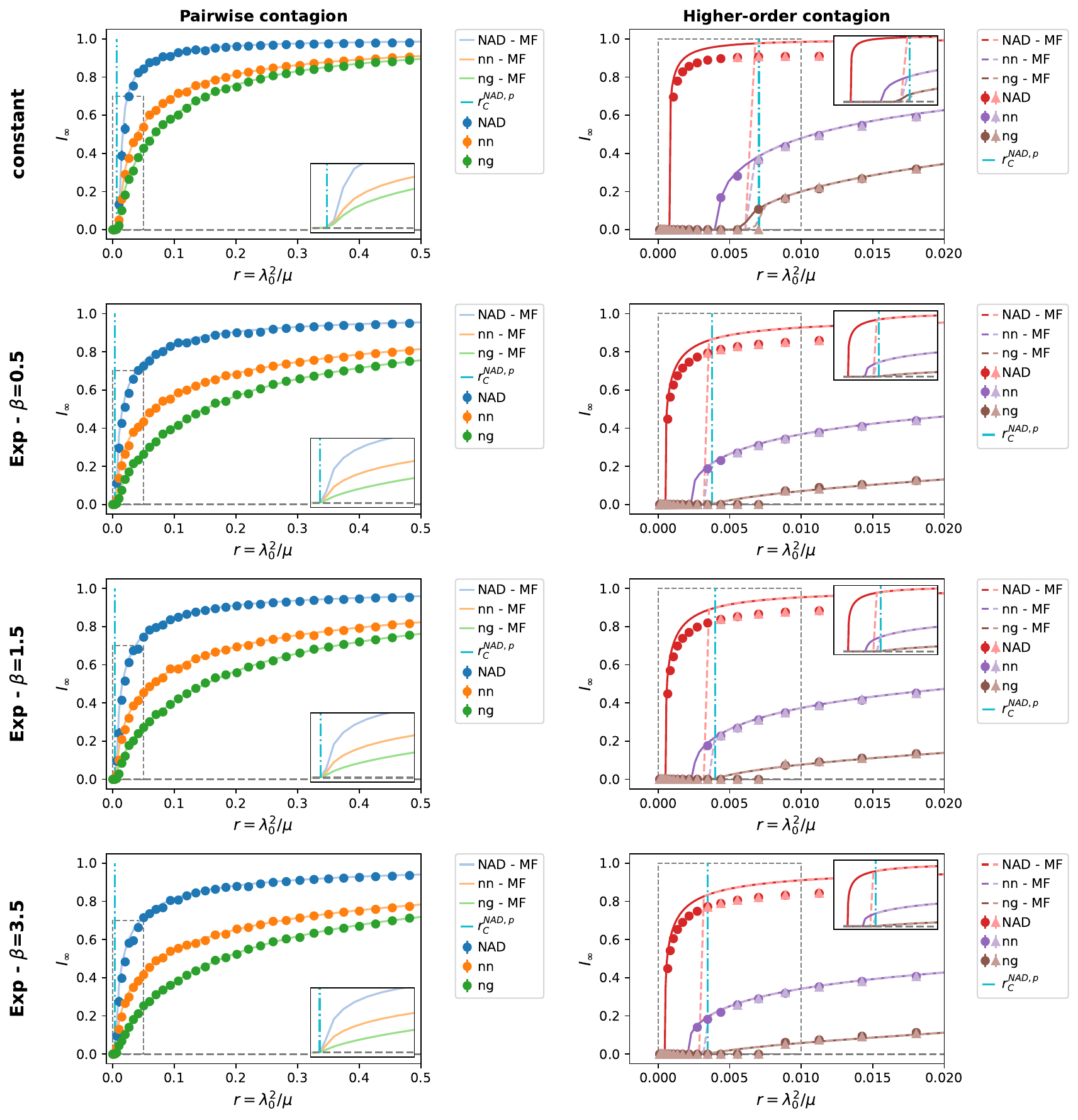}
\caption{\textbf{Phase diagram for the pairwise and higher-order contagion processes - Synthetic - I.} Each row corresponds to a different synthetic hypergraph generated (see labels): for each of them we show the epidemic prevalence in the asymptotic steady state $I_{\infty}$ as a function of the effective infection rate $r$, for the non-adaptive case (NAD) and for the two adaptation strategies ($nn$, $ng$), respectively for the pairwise (first column) and higher-order contagion process (second column). For each case, we show the results of numerical simulations (markers) and of integration of the mean-field equations (lines). For the higher-order contagion case, we consider simulations and numerical integration starting with a low (triangles and dashed lines) and a high (dots and solid lines) fraction of infected nodes. In both panels, the simulation results are averaged over $200$ simulations, $\nu=4$, $\theta=0.3$, the inset represents a zoom on the mean-field results in the area marked by a dashed rectangle and the vertical dash-dotted line indicate $r_C^{NAD,p}$. For numerical simulations, error-bars correspond to one standard deviation (sometimes they are not visible since smaller than the corresponding markers).}
\label{fig:figure9}
\end{figure*}

\begin{figure*}[ht!]
\includegraphics[width=\textwidth]{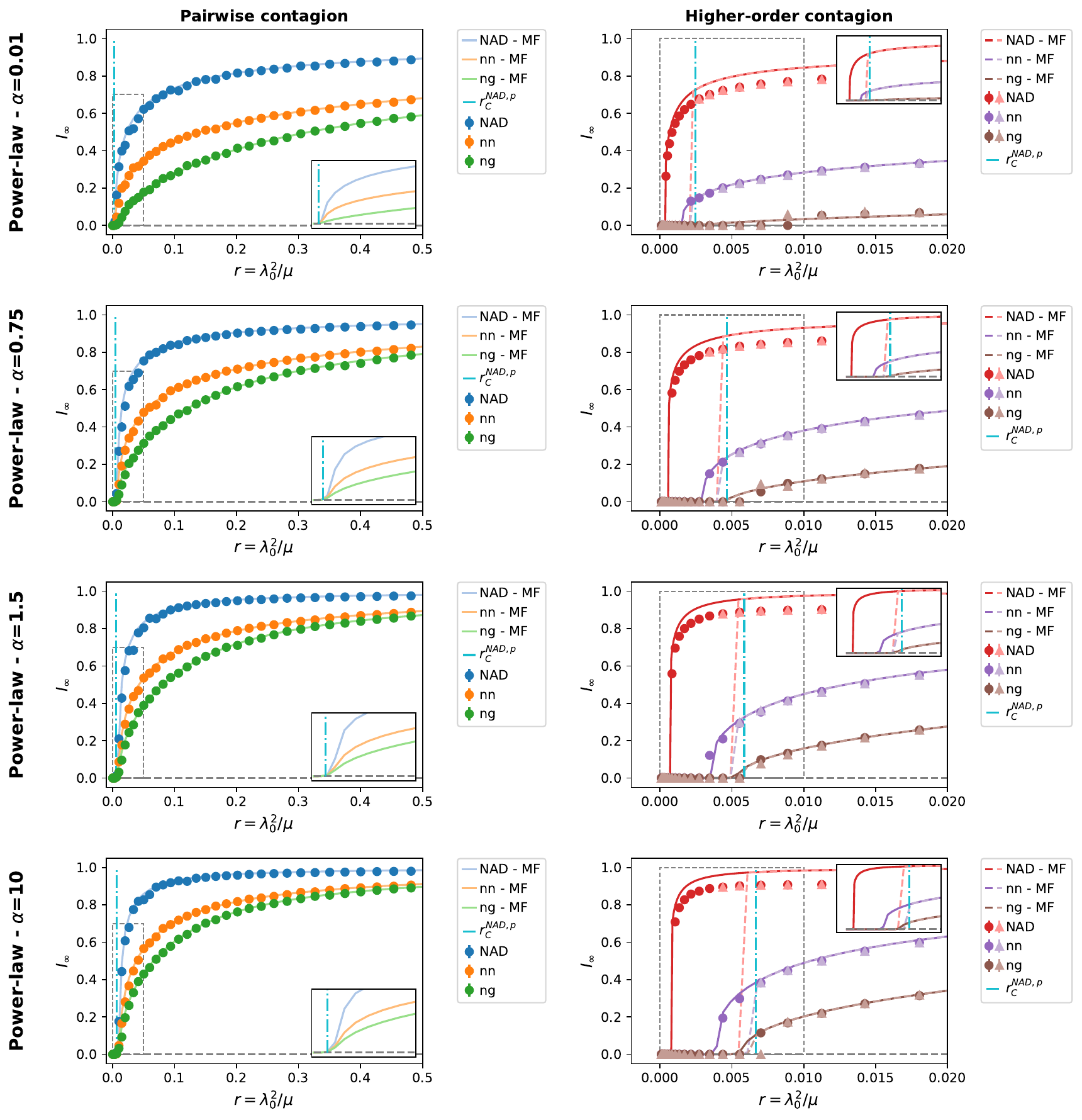}
\caption{\textbf{Phase diagram for the pairwise and higher-order contagion processes - Synthetic - II.} Each row corresponds to a different synthetic hypergraph generated (see labels): for each of them we show the epidemic prevalence in the asymptotic steady state $I_{\infty}$ as a function of the effective infection rate $r$, for the non-adaptive case (NAD) and for the two adaptation strategies ($nn$, $ng$), respectively for the pairwise (first column) and higher-order contagion process (second column). For each case, we show the results of numerical simulations (markers) and of integration of the mean-field equations (lines). For the higher-order contagion case, we consider simulations and numerical integration starting with a low (triangles and dashed lines) and a high (dots and solid lines) fraction of infected nodes. In both panels, the simulation results are averaged over $200$ simulations, $\nu=4$, $\theta=0.3$, the inset represents a zoom on the mean-field results in the area marked by a dashed rectangle and the vertical dash-dotted line indicate $r_C^{NAD,p}$. For numerical simulations, error-bars correspond to one standard deviation (sometimes they are not visible since smaller than the corresponding markers).}
\label{fig:figure10}
\end{figure*}

\begin{figure*}[ht!]
\includegraphics[width=\textwidth]{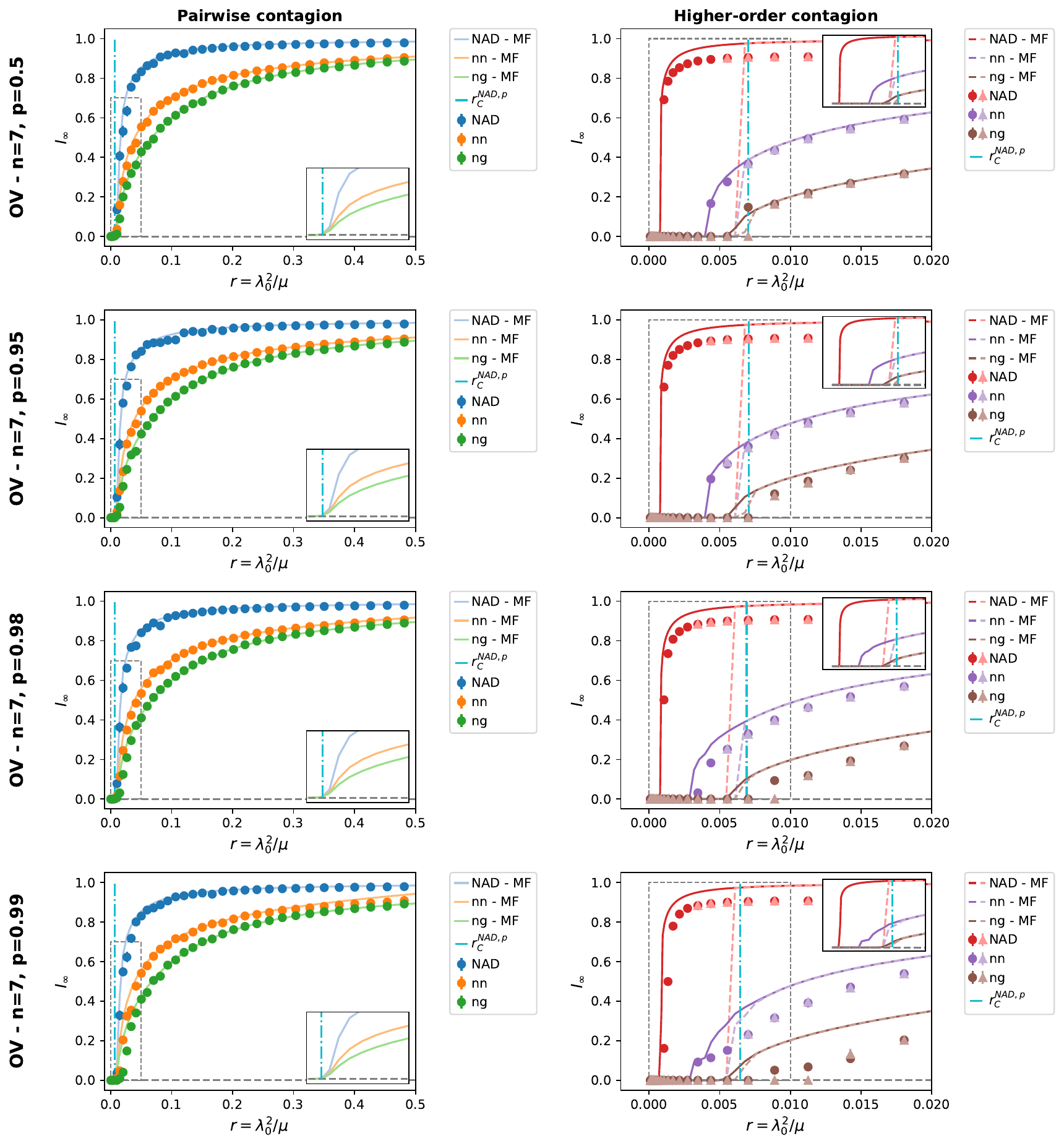}
\caption{\textbf{Phase diagram for the pairwise and higher-order contagion processes - Synthetic - III.} Each row corresponds to a different synthetic hypergraph generated (see labels): for each of them we show the epidemic prevalence in the asymptotic steady state $I_{\infty}$ as a function of the effective infection rate $r$, for the non-adaptive case (NAD) and for the two adaptation strategies ($nn$, $ng$), respectively for the pairwise (first column) and higher-order contagion process (second column). For each case, we show the results of numerical simulations (markers) and of integration of the mean-field equations (lines). For the higher-order contagion case, we consider simulations and numerical integration starting with a low (triangles and dashed lines) and a high (dots and solid lines) fraction of infected nodes. In both panels, the simulation results are averaged over $200$ simulations, $\nu=4$, $\theta=0.3$, the inset represents a zoom on the mean-field results in the area marked by a dashed rectangle and the vertical dash-dotted line indicate $r_C^{NAD,p}$. For numerical simulations, error-bars correspond to one standard deviation (sometimes they are not visible since smaller than the corresponding markers).}
\label{fig:figure11}
\end{figure*}

Supplementary Fig.s \ref{fig:figure9}-\ref{fig:figure11} show the epidemic phase diagram for the different adaptive strategies, for the pairwise and higher-order contagion processes, on the synthetically generated hypergraphs. These results show that the heterogeneity of the hyperdegree distribution plays an important role in both the pairwise and higher-order contagion processes, influencing the epidemic threshold, the phase transition, and the active phase, while the overlap seems to have a more limited impact. First, the more heterogeneous the underlying hypergraph, the greater the differences between the $nn$ and $ng$ strategies in the active phase and their effectiveness compared to the NAD case: we explore this aspect in detail in \cite{mancastroppa2025}.\\

\begin{figure*}[ht!]
\includegraphics[width=\textwidth]{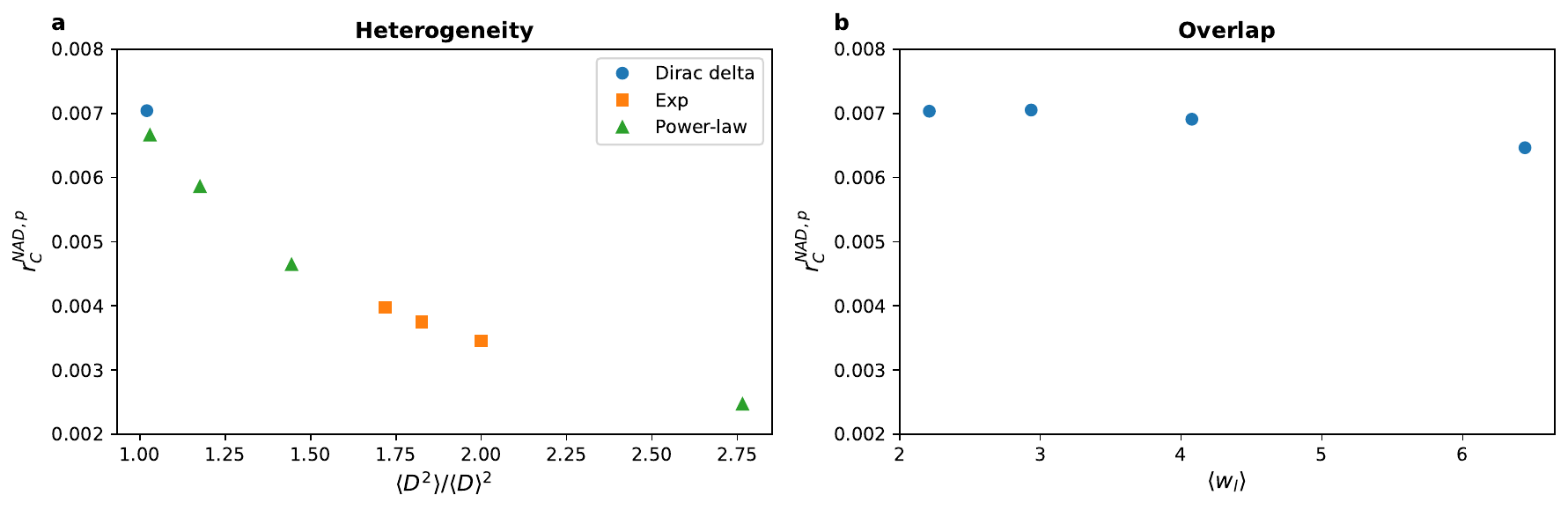}
\caption{\textbf{Effect of hyperdegree heterogeneity and hyperedges overlap on pairwise contagion process.} In panel \textbf{a} we show the epidemic threshold $r_C^{NAD,p}$ of the pairwise contagion process as a function of the heterogeneity $\langle D^2 \rangle/\langle D \rangle^2$ of the hyperdegree distribution. The epidemic threshold is obtained considering its mean-field analytical estimation; each marker correspond to a specific synthetic hypergraph of the PD method (different markers identify different PD models -see legend-, while markers with the same shape and color correspond to different model parameters). In panel \textbf{b} we show the epidemic threshold $r_C^{NAD,p}$ as a function of the hyperedge overlap $\langle w_l \rangle$; each marker correspond to a specific synthetic hypergraph of the OV method obtained with a different $p$ value (while $n=7$ is fixed).}
\label{fig:figure12}
\end{figure*}

\begin{figure*}[ht!]
\includegraphics[width=\textwidth]{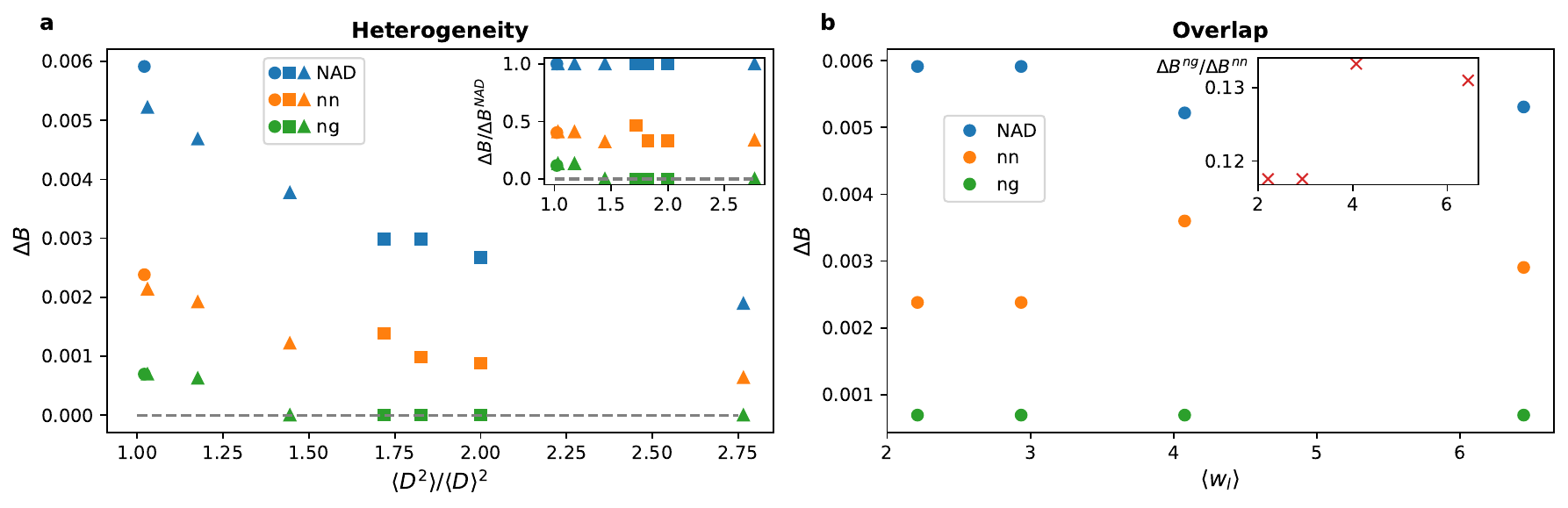}
\caption{\textbf{Effect of hyperdegree heterogeneity and hyperedges overlap on higher-order contagion process.} In panel \textbf{a} we show the width of the bistability region $\Delta B = r_C^{NAD,p}-r_C^{discontinuous}$ of the higher-order contagion process as a function of the heterogeneity $\langle D^2 \rangle/\langle D \rangle^2$ of the hyperdegree distribution. $\Delta B$ is obtained considering the results of the numerical integration of the mean-field equations; each marker correspond to a specific synthetic hypergraph of the PD method (different markers identify different PD models -see legend of Seupplementary Fig. \ref{fig:figure12}- while markers with the same shape and color correspond to different model parameters). We consider $\Delta B$ for both the NAD case and for the two adaptive strategies $nn$ and $ng$ (see colors). In the inset we report the ratio $\Delta B/\Delta B^{NAD}$ for the different strategies, as a function of $\langle D^2 \rangle/\langle D \rangle^2$. In panel \textbf{b} we show the width of the bistability region $\Delta B$ as a function of the hyperedge overlap $\langle w_l \rangle$; each marker correspond to a specific synthetic hypergraphs of the OV method obtained with a different $p$ value (while $n=7$ is fixed). We consider $\Delta B$ for both the NAD case and for the two adaptive strategies $nn$ and $ng$ (see colors). In the inset we report the ratio $\Delta B^{ng}/\Delta B^{nn}$ as a function of $\langle w_l \rangle$}
\label{fig:figure13}
\end{figure*}

Focusing on the phase transition, Supplementary Fig. \ref{fig:figure12} shows the impact of hyperdegree heterogeneity and hyperedges overlap on the mean-field epidemic threshold of the pairwise contagion processes, considering the different hypergraphs generated synthetically. Analogously, Supplementary Fig. \ref{fig:figure13} shows the impact of hyperdegree heterogeneity and hyperedges overlap on the width of the bistability region for the higher-order contagion process, considering the different hypergraphs generated synthetically. In general, heterogeneity strongly reduces the epidemic threshold for pairwise contagion, as a typical effect of topological heterogeneities \cite{satorras2015epidemic}, favoring the epidemic; the overlap also slightly reduces the epidemic threshold, but with a very limited effect. In the case of higher-order contagion, heterogeneity reduces the width of the bistability region and amplifies the differences between the different strategies and their impact, by favoring the disappearance of the bistability region. The hyperedges overlap, on the other hand, has a limited effect, which appears to primarily influence the difference in performance of the two adaptive strategies \cite{mancastroppa2025}.

\newpage

\newpage

\newpage

\newpage

\newpage

\newpage

\newpage

\clearpage
\bibliographystyle{naturemag}
\bibliography{references_HO_adaptive_epi_supp}